\documentstyle[prl,aps,twoside,psfig]{revtex}

\begin{document}

\draft

\title{Collective modes in uniaxial incommensurate-commensurate systems\\
 with the real order parameter}

\author{V. Danani\'{c}}  
 \address{ Department of Physics, Faculty of Chemical Engineering and 
 Technology,  University of Zagreb\\
 Maruli\'{c}ev trg 19, 10000 Zagreb, Croatia}
\author{A. Bjeli\v{s} and M. Latkovi\'{c}}
 \address{Department of Theoretical Physics, Faculty of Science, University 
 of Zagreb\\
 Bijeni\v{c}ka 32, 10000 Zagreb, Croatia}

\maketitle

\begin{abstract}

The basic Landau model for uniaxial systems of the II class is nonintegrable,
and allows for various stable and metastable periodic configurations, beside
that representing the uniform (or dimerized) ordering. In the present paper 
we complete the analysis of this model by performing the second order 
variational procedure, and formulating the combined Floquet-Bloch approach 
to the ensuing nonstandard linear eigenvalue problem. This approach enables 
an analytic derivation of some general conclusions on the stability of 
particular states, and on the nature of accompanied collective excitations. 
Furthermore, we calculate numerically the spectra of collective 
modes for all states participating in the phase diagram, and analyze 
critical properties of Goldstone modes at all second order and 
first order transitions between disordered, uniform and periodic states. 
In particular it is shown that the Goldstone mode softens as the underlying
soliton lattice becomes more and more dilute.  

\end{abstract}

\pacs{05.10.-a, 05.70.Fh, 64.70.Rh}

\section{Introduction} 

One of the most useful insights into the properties of stable and metastable
ordered states in many body systems follows from the investigations of 
accompanying collective
modes, excitations with a coherent participation of (semi)macroscopic number 
of particles. The attention is usually focused on the lowest branch in the
spectrum. If it is of Goldstone  type, i. e. gapless (e. g. acoustic) in 
the long wavelength limit ($k \rightarrow 0$), there is a continuous 
degeneracy in the characterization of ordered state, associated with the 
breaking of symmetry of the high temperature thermodynamic phase. Without 
the continuity in the degeneracy one has instead a finite gap at $k = 0$.

Obvious extrinsic causes for the gap in the Goldstone mode are impurities,
defects in the crystal structure, etc. Another cause for
the gap is the presence of long range interactions~\cite{lra}.
We do not consider either of these mechanisms here, but remind that, 
as is well known in charge density wave materials~\cite{cdw,evo}, they
may play a decisive role in the collective dynamics of ordered state.  
Instead, we concentrate on the systems in which the above distinction 
regarding the degeneracy of ordered state(s) has its origin in short 
range interactions. Those are numerous materials that show one or more types 
of uniaxially modulated orderings with periodicities which may be 
commensurate or incommensurate with respect to the underlying crystal 
lattice~\cite{blinc,cumm}. 

Let us at the beginning invoke some simple widely accepted conclusions,
accumulated through intense theoretical and experimental investigations on 
these incommensurate-commensurate (IC) systems in last few decades. In an 
ideal case of sinusoidal modulation the spectrum of collective excitations 
contains two types of modes, phasons and amplitudons, representing 
linearized fluctuations of phase and amplitude of the 
order parameter respectively. While the amplitudon mode has a finite gap 
below the critical temperature, the phason mode is acoustic if the free 
energy of corresponding state does not depend on the relative phase of 
ordered modulation and crystal lattice. In other words one has the
continuous degeneracy with respect to this relative phase. It is strictly
fulfilled only if the modulation is incommensurate with respect to the 
periodicity of crystal lattice. 

For commensurate modulations the free energy depends on the relative phase, 
as is easily seen already from the standard Landau expansions in which the 
lattice discreteness is taken into account by keeping a leading Umklapp 
contribution. Within this standard and frequently explored
model~\cite{mcmil,toledano}, which leads 
to the simple variational equation of sine-Gordon type, the phason mode 
acquires a gap which is finite only for the strict commensurate ordering, and 
diminishes rapidly (exponentially) as the order of commensurability increases.  
For other modulations, which may have the form of dilute soliton lattices, the
Goldstone mode remains gapless, although among these modulations there are
solutions with commensurate periodicities close to the exempted leading
commensurability. In other words, within this model one does not distinguish 
between "secondary" commensurate orderings and incommensurate orderings.   
This is the consequence of a crude simplification made by retaining only 
one Umklapp term in the free energy. The recent analysis shows that already
after taking into account two leading Umklapp terms the phase diagram becomes 
qualitatively different~\cite{lbpl,lbprb}. It contains a finite number of 
commensurate states, and shows a harmless staircase, i. e. a series of first 
order transitions between neighboring states. The Goldstone mode is then 
expected to have a finite gap for each state participating in the phase 
diagram. 

The Landau models for the orderings with spatial modulations are generally
justified providing the interactions responsible for their stabilization 
are weak enough, so that the variations of order parameter (defined 
with respect to the appropriately chosen star of wave vectors) are slow at the 
scale of lattice constant. Two crucial simplifications are then allowed, 
namely the gradient expansion and the perturbative treatment of lattice 
discreteness through the truncation of the sum of Umklapp contributions. 

In the opposite regime of strong couplings the above spatial continuation 
is not allowed, and the lattice discreteness leads to qualitatively different
properties of phase diagrams and related spectra of excitations, established 
by numerous analytical, and particularly numerical, studies of spin (e.g. 
Ising) \cite{bakbo,selke}, displacive (e.g. Frenkel-Kontorova) 
\cite{fkaub}, and electron-phonon (e.g. Holstein) \cite{holaub,lorthe} 
discrete models. Characteristically for such models, either a finite, 
sometimes large, number of commensurate modulations in the cases of harmless 
staircase, or the infinity of them in the cases of complete devil's staircases, 
can participate in the phase diagram. All commensurate states then have 
lowest branches of collective excitations with finite gaps in the limit
$k\rightarrow 0$. We repeat that, while none of these possibilities can be 
reproduced by Landau model with one Umklapp term, the former harmless 
staircases with a finite number of commensurate states are realized already 
within extended Landau models with only two Umklapp terms taken into 
account~\cite{lbpl,lbprb}.   

The analysis of Frenkel-Kontorova and Holstein models established also a new 
type of instability that involves incommensurate modulations, the so-called
transition by breaking of analyticity~\cite{aubry}. Namely, by increasing the 
coupling constant~\cite{aq}, or by decreasing temperature
~\cite{baesens,lorenzo}, the smooth envelope of an incommensurate periodic 
modulation becomes nonanalytic. The free energy then depends nonanalytically 
on the relative phase of modulation and underlying lattice. As a consequence a 
finite gap opens in the Goldstone branch of collective excitations even for 
incommensurate modulations. 

Already from the beginning of investigations on discrete models it was 
realized that the above complex features in phase diagrams and spectra of 
collective excitations have their origin in the nonintegrability of these 
models, i. e. in the nontrivial chaotic structures of corresponding phase 
spaces. In this respect it is important to emphasize that, either in their 
basic form or after the inclusion of further terms, Landau free energy 
expansions are as a rule the examples of nonintegrable functionals. 
For example, while the sine-Gordon model with one Umklapp term, as a 
basic model for class I of IC systems, is integrable, already the inclusion 
of another Umklapp term brings in the nonintegrability~\cite{lbpl,lbprb}. 

The situation is even more intriguing for the class II, i. e. for IC materials 
with modulations having the period close or equal either to the original or to 
the dimerized unit cell of crystal lattice. There are numerical~\cite{dbpra} 
and analytical~\cite{dbpre} indications that already the
minimal~\cite{hornreich,michelson}, as well as slightly extended~\cite{is1},
models for this class are not integrable. The consequences of this
nonintegrability on the phase diagram are discussed in detail in 
Ref.~\cite{dbpre}. In particular, it is shown that, in addition to simple 
disordered, commensurate [i. e. (anti)ferro] and (almost) sinusoidal 
incommensurate states, included into previous 
analyses~\cite{michelson,bruce1,is1}, the phase diagram contains also an 
enumerable family of metastable solutions with the periodic alternations of 
commensurate and incommensurate sinusoidal domains. 

In the present work we calculate the spectrum of collective modes for 
stable and metastable states in systems of class II. The corresponding
Landau model is particularly convenient for the discussion of questions 
raised in this Introduction, since it is nonintegrable, and, in addition,
the accompanying phase diagram comprises both commensurate and incommensurate 
(meta)stable states. Our main aim is to investigate to what extent are the 
collective modes influenced by the nonintegrability of, here continuous, 
free energy functional. Furthermore, by analyzing Goldstone modes for the 
modulated states of the model under consideration~\cite{michelson,is1} we also 
resolve some controversies present in the 
literature~\cite{ls,aramburu,mashiyama,san,ss} on its applicability in the 
description of incommensurate phases in systems of class II. The equivalent 
analysis for the class I, i. e. for the Landau model with two Umklapp terms, 
will be presented elsewhere~\cite{ldbun}.

The plan of the paper is as follows. The free energy functional for the 
class II is introduced in Sec. II. In Sec. III we perform the variational 
procedure up to the second order, taking care about some specific questions 
related to the 
thermodynamic minimization~\cite{dbprl}. The linear eigenvalue problem
associated to the second order variational procedure is discussed in Sec. IV.   
Here we encounter a generalized Hill problem, since the systems includes four 
coupled first order equations (in contrast to the standard cases with two 
equations), and furthermore, since we are looking for the collective modes 
of highly multiharmonic periodic states. We therefore do not follow a standard 
way, appropriate for simple sinusoidal incommensurate orderings, but develop 
for the first time a general formalism, applicable also to other types 
of Landau models. This formalism enables the determination of Floquet 
exponents, and of corresponding Bloch basis of eigenfunctions which we consider 
in Sec.V. The numerical results for the collective modes of all (meta)stable 
states appearing in the phase diagram are presented in Sec. VI. Concluding 
remarks along the lines specified in the previous paragraph are given in 
Sec. VII.

\section{Model}

The free energy functional for the uniaxial incommensurate systems of  
class II is given by
\begin{equation}
 \tilde f[\tilde u]=\frac{1}{2\tilde L}\int_{- \tilde L}^{\tilde L}
 \left[d\left(\frac{d^{2}\tilde u}{d\tilde z^{2}}
 \right)^{2} +
 c\left(\frac{d\tilde u}{d\tilde z}\right)^{2} + a\tilde u^{2} + 
 {\scriptstyle \frac{1}{2}} b\tilde u^{4}\right]d\tilde z
\label{fe1}
\end{equation}
where $\tilde u$ represents the real order parameter and $\tilde L$ is the 
length of the system. This functional is the simplest (minimal) Landau 
expansion for the systems with minima of free energy density in the 
reciprocal space close to the center of Brillouin zone, or to the part of 
its border perpendicular to the uniaxial direction. Then $c < 0$, and one has 
to add a highly nontrivial term with the second derivative of $\tilde u$ (and 
presumably positive coefficient $d$) in order to ensure the boundness of  
Landau expansion in the 
reciprocal space. The rest of the expansion (\ref{fe1}) is standard, with 
$b > 0$, and $a$ becoming negative below the critical temperature of the 
transition from disordered to uniform (ferro) or dimerized (antiferro) 
phase. We limit the further analysis to the most interesting regime
characterized by $c < 0$. It includes the incommensurate ordering and 
the transition to the commensurate ordering (but does not include the 
transition from disordered to commensurate state which takes place for 
$c > 0$) \cite{hornreich,michelson}. In this regime the useful dimensionless 
quantities are
\begin{equation}
z=\sqrt{-\frac{c}{d}}\tilde z,\,\,\,\ L=\sqrt{-\frac{c}{d}}\tilde L,\,\,\,\ 
u(z)= - \frac{\sqrt{bd}}{c} \tilde u(\tilde z),\,\,\,\ f[u]=\frac{bd^2}{c^4}
\tilde f[\tilde u].
\label{red}
\end{equation}
The model (\ref{fe1}) can be now represented as the one-parameter problem,
\begin{equation} 
f[u]=\frac{1}{2L}\int_{-L}^{L}\left[\left(\frac{d^{2}
 u}{dz^{2}}\right)^{2}-\left(\frac{du}{dz}
 \right)^{2}+\lambda\,u^{2}+{\scriptstyle\frac{1}{2}}u^{4}\right]dz,
\label{fe2}
\end{equation}
with $\lambda \equiv ad/c^2$. The parameterization of the phase 
diagram in the regime $c < 0$ is thus very simple, since all relationships 
between different (meta)stable states (like phase transitions, ranges of 
coexistence of two or more states, etc) can be presented in the 
one-dimensional $\lambda$-space. The knowledge of the actual dependence of 
this parameter, as well as of the scales which enter into the reduced 
quantities (\ref{red}), on the original physical parameters, in particular on 
temperature, goes together with the specification of microscopic background 
behind the phenomenological free energy (\ref{fe1}). This is a necessary step 
in any comparison of phase diagram for the model (\ref{fe2}) with experimental 
data for a given material. 

In the previous works~\cite{dbpra,dbpre} on the functional (\ref{fe2}) we 
have determined thermodynamically stable states, i. e. its local minima, 
without taking into considerations statistical fluctuations outside these 
minima. This mean-field type of approximation is inappropriate for 
(quasi) one-dimensional systems. It is however usually sufficient for 
three-dimensional uniaxial systems with strong enough couplings in the 
perpendicular directions, on which we concentrate here. 

The thermodynamic extremalization of functional (\ref{fe2}) consists of 
the standard variational procedure that is equivalent to the classical 
mechanical one and leads to the corresponding Euler-Lagrange (EL) equation
\begin{equation} 
\frac{d^{4}u}{dz^{4}}+\,\frac{d^{2}u}{dz^{2}}+\lambda\,u+u^{3}=0,
\label{el}
\end{equation}
and of the extremalization that involves boundary conditions or some equivalent 
set of parameters. The general procedure that carefully takes 
into account the latter aspect is proposed in Ref.~\cite{dbprl}. The most 
interesting result of this approach is obtained for the functional with the
kernel that is not explicitly $z$-dependent. Then the relation
\begin{equation} 
F = -H, 
\label{condA}
\end{equation} 
holds for each thermodynamic extremum $u_0(z)$. Here $F$ is the corresponding 
averaged free energy, and $H$ is the integral constant of the problem (\ref{el})
which corresponds to the Hamiltonian in classical mechanics.

Early considerations of the model (\ref{fe1}) led to the suggestion that the
mean-field phase diagram~\cite{hornreich,michelson,bruce1,bruce2,horn2} 
contains only disordered [$u_d(z) = 0$], commensurate [$u_c(z) = 
\pm\sqrt{-\lambda}$], and (almost) sinusoidal [$u_s(z) \approx 2/\sqrt{3}
 (\sqrt{1/4 - \lambda})\sin{(z/\sqrt{2})}$] incommensurate 
orderings. The commensurate state is thermodynamically stable for 
$\lambda < -1/8$ (and for $a < 0$ in the range $c > 0$). The incommensurate 
state is stable in the range $-2 < \lambda < \lambda_{id} = 1/4$, while the 
first order phase transition between the 
commensurate and the incommensurate states occurs at $\lambda_{ic} = -1.112$. 
The more precise values, obtained after taking into account 
corrections from higher harmonics in the sinusoidal ordering~\cite{dbpre}, are
$-1.835 < \lambda < \lambda_{id}$ and $\lambda_{ic} = -1.177$. Also, the wave 
number of this ordering, $q$, slightly deviates from $1/\sqrt{2}$ [i. e. 
from $\sqrt{-c/(2d)}$ in the original parameters of Eq.~(\ref{fe1})] as 
one approaches the left edge of instability, $\lambda \rightarrow -1.835$.  

While by above solutions of EL equation~(\ref{el}) one exhausts all absolute
minima of the free energy~(\ref{fe1}), the more involved numerical
analysis~\cite{dbpra,dbpre} showed the existence of an enumerable series of 
periodic solutions which are metastable in finite ranges of the parameter 
$\lambda$. The corresponding phase diagram is shown in Fig.~\ref{phdi} in which 
we ascribe to various solutions symbolic words introduced in Ref.~\cite{dbpre}.
By their physical content the metastable solutions from Fig.~\ref{phdi}
represent periodic trains of successive sinusoidal and uniform segments
(see Fig.~1 in~\cite{dbpre}), and, as domain patterns, complete in a natural 
way, as an inherent outcome of nonintegrable model~(\ref{fe1}), the phase 
diagram in the range of coexistence of two corresponding basic types of 
orderings.

\section{Second order variational procedure}

The question on which we concentrate now is the thermodynamic stability of a 
given state $u(z)$ which is a solution of EL equation~(\ref{el}) and 
fulfills additional conditions of Ref.~\cite{dbprl}. To this end we have to go 
beyond the linear terms in the extremalization procedure. Let us therefore at 
first extend the standard variational procedure to the second order. Later on 
we shall shortly consider the conditions which follow from the minimization of 
boundary conditions.
 
Let $\eta(z)$ be the infinitesimal variation with respect to $u(z)$, obeying
usual conditions at the boundaries $z=0$ and $z=L$, 
\begin{equation} 
\eta(z=0) = \eta(z=L) =\eta'(z=0) =\eta'(z=L) = 0. 
\label{bv}
\end{equation}
After performing standard partial integrations, 
\begin{eqnarray} 
\frac{1}{L}\int_{0}^{L} (\eta')^2 dz & = &
\frac{1}{L}\eta' \eta \arrowvert_{0}^{L} - 
\frac{1}{L}\int_{0}^{L} \eta''\eta dz , \nonumber \\
\frac{1}{L}\int_{0}^{L} (\eta'')^2 dz & = &
\frac{1}{L}\eta'' \eta' \arrowvert_{0}^{L} 
- \frac{1}{L}\eta'''\eta \arrowvert_{0}^{L} + 
\frac{1}{L}\int_{0}^{L} \eta^{IV}\eta dz,
\label{parint}
\end{eqnarray}
the quadratic contribution to the corresponding variation of free energy
functional~(\ref{fe2}) can be expressed in the form
\begin{equation} 
\delta^{2}f \equiv f[u + \eta] - f[u] = \frac{1}{L}\int_{0}^{L} dz \eta 
{\cal D} \eta,
\label{fevar}
\end{equation}
with 
\begin{equation} 
{\cal D} \equiv \frac{d^4}{dz^4} + \frac{d^2}{dz^2} + \lambda + 3 u^{2}.
\label{calD}
\end{equation}
The linear differential operator (\ref{calD}) defines the eigenvalue problem
\begin{equation} 
{\cal D}\eta _{\Lambda}\equiv \eta''''_{\Lambda}(z)
+\eta''_{\Lambda}(z)+\left[\lambda+3u^{2}(z)\right]\eta_{\Lambda}(z) = 
\Lambda\eta_{\Lambda}(z),
\label{eta}
\end{equation}
with the boundary conditions for $\eta(z)$ specified by Eq.~(\ref{bv}).
The necessary condition for the thermodynamic stability of solution
$u(z)$ is given by the requirement that the spectrum $\Lambda$ is
non-negative for all normalizable solutions $\eta_{\Lambda}(z)$ of 
the problem~(\ref{calD},\ref{bv}).

Since the above procedure strictly respects the boundary conditions~(\ref{bv}),
it is entirely  equivalent to that usually used in classical 
mechanics. As a consequence the obtained condition for the stability of 
a given solution $u(z)$ holds for any value of the sample length $L$. 
However, neither the extremal solution $u(z)$ of EL equation~(\ref{el}), nor 
the corresponding conditions of thermodynamic stability, should be sensitive 
to the conditions imposed on the sample surfaces in the physically relevant 
thermodynamic limit $L\rightarrow\infty$. Therefore the stability condition 
can be generalized in this limit. In particular, we may ignore the boundary 
conditions~(\ref{bv}), and perform the variational procedure for any 
infinitesimal variation $\eta(z)$, noting, for later purposes, that the 
requirement of infinitesimality excludes variations $\eta(z)$ which would 
scale as $|z|^\beta$ with $\beta > 0$. 

Performing the same steps as before, 
but now with neglected surface terms in Eqs.~(\ref{parint}) (which scale as 
$1/L$), we come again to the linear eigenvalue problem~(\ref{eta}), but without 
a specification on boundary conditions. This means that {\em any} complete 
set of eigenfunctions $\eta_{\Lambda}(z)$ with the eigenvalues $\Lambda$
[which are themselves characterized solely by the linear differential
equation~(\ref{eta})] can be used in the representation of a given variation 
$\eta(z)$, and in the corresponding diagonal representation of the free energy 
(\ref{fevar}). The condition ${\Lambda}\geq 0$ thus guarantees the 
thermodynamic stability of the solution $u(z)$ with respect to {\em any} 
infinitesimal variation, provided the system has the well-defined thermodynamic 
limit as specified above. We note that by this relaxation of boundary 
conditions~(\ref{bv}) we extend the standard "classical mechanical"
second order variational procedure by including a part of, but still not 
all, "thermodynamic" variations. The discussion of this question in Appendix A
suggests that the criterion of thermodynamical stability is probably entirely
covered by the eigenvalue problem~(\ref{eta}). 

The concise definition of the thermodynamical stability, 
i. e. of the stability of any (absolutely stable or metastable) local minimum 
of thermodynamic functional with respect to small fluctuations, is thus:
\begin{itemize}
\item A given solution of EL equation (\ref{el}) is thermodynamically stable if 
and only if {\em all} solutions of Eq.(\ref{eta}) for {\em any} $\Lambda<0$ are 
non-normalizable.
\end{itemize}
For later purpose it is appropriate to introduce here also the concept of 
orbital stability, relevant for the behavior of particular solutions in the 
phase space:
\begin{itemize}
\item A given solution $u(z)$ of EL equation (\ref{el}) is orbitally stable if
and only if {\em all} solutions of Eq.(\ref{eta}) for $\Lambda=0$ are 
normalizable.
\end{itemize}

In the next section the above definitions will be used in the study
of stability of homogeneous and periodic configurations $u(z)$. The crucial
assumption in this respect is that the solutions of Eq.(\ref{eta}) smoothly 
depend on both parameters $\lambda$ and $\Lambda$. 

Before embarking into the calculation of the spectrum of eigenvalue 
problem~(\ref{eta}), we invoke its general property which follows from the 
fact that the density of free energy functional~(\ref{fe2}) does not 
depend explicitly on the spatial coordinate $z$.
Then there exists a normalizable solution of equation~(\ref{eta}) with 
$\Lambda=0$, namely $\eta_0(z)\propto u'(z)$. This is the Goldstone mode 
that follows from the translational invariance of free energy 
functional~(\ref{fe2}), by which $u(z+z_{0})$ with arbitrary $z_{0}$ is the 
solution of EL equation~(\ref{el}) if $u(z)$ is its solution.
We note that Eq.(\ref{eta}) then has, together with the above Goldstone 
mode, another solution of the form $\eta_{1}(z)=w(z)+z\cdot u'(z)$, where 
$w(z)$ is some periodic function of the same period as that of the Goldstone 
mode. Although $\eta_{1}(z)$ is non-normalizable {\em i.e.} its norm 
grows as a power of $L$, we consider this non-normalizability as {\em marginal}.
The power law growth of a solution of Eq.(\ref{eta}) is much easier to control 
than possible exponential growth of the remaining solutions, if there are any. 
For special values of the parameter $\lambda$ figuring in Eq.(\ref{eta}) with 
$\Lambda=0$ the only normalizable solution is the Goldstone mode $u'(z)$, while 
other solutions have a power law growth in $z$, $z^{n}u'(z)$ with $n \leq 3$. 
These special values of $\lambda$ denote the edges of thermodynamical 
metastability of the corresponding configuration $u(z)$. 

\section{Floquet theory}

The analysis of the eigenvalue problem~(\ref{eta}) with periodic functions
$u(z)$ is based on general Floquet and Bloch theorems for linear 
differential equations with periodic coefficients. It will be performed in two
stages, covered by this and the next section. The aim of the first one, based 
on the Floquet's approach, is to answer the question: {\em Whether there exists 
a normalizable solution $\eta_{\Lambda}(z)$ for a given value of $\Lambda$}? In 
the second stage we calculate the set of values $\Lambda$ for which 
normalizable solutions exist, i.e. the spectrum of collective modes, by using 
the Bloch's wave number representation.

We begin by showing that the set of values of $\Lambda$ for which the 
corresponding normalizable solutions $\eta_{\Lambda}(z)$ may exist is
bounded from below. To this end let us rewrite Eq.~(\ref{eta}) in the form 
\begin{equation}
 {\tilde{\cal D}}^{2}\eta_{\Lambda}(z)+3\,u(z)^{2}\eta_{\Lambda}(z)=
 \left(\Lambda+\frac{1}{4}-\lambda\right)\eta_{\Lambda}(z)\,\,\,,\,\,\,
{\tilde{\cal D}}\equiv\frac{d^{2}}{dz^{2}}+\frac{1}{2},
\label{etah}
\end{equation}
and introduce the norm of the function $\eta_{\Lambda}(z)$,
\begin{equation}
\parallel \eta_{\Lambda}\parallel^{2}
\equiv\left<\eta_{\Lambda}^{*}\eta_{\Lambda}\right> =
\frac{1}{L}\int_{0}^{L}\eta_{\Lambda}(z)^{*}\eta_{\Lambda}(z)dz.
\label{norm}
\end{equation}
After multiplying Eq.~(\ref{etah}) by $\eta_{\Lambda}^{*}(z)$ and integrating 
with respect to $z$ we get
\begin{equation}  
\parallel {\tilde{\cal D}}\eta_{\Lambda}(z) \parallel^{2}+
3\, \parallel u(z)\eta_{\Lambda}(z) \parallel^{2}= \left(\Lambda+
\frac{1}{4}-\lambda\right)\parallel \eta_{\Lambda}(z)\parallel^{2} .
\label{etaint}
\end{equation}
Here it is taken into account that the operator ${\tilde{\cal D}}$ is 
hermitean and the function $u(z)$ is real. Since the left-hand side of 
Eq.~(\ref{etaint}) is strictly positive, we conclude that 
\begin{equation}
\Lambda\geq \Lambda_{min}= \lambda -\frac{1}{4}
\label{minLambda}
\end{equation}
for each $\Lambda$ for which the the norm~(\ref{norm}) of the function
$\eta_{\Lambda}(z)$ exists. In particular, this means that it is sufficient to 
reduce a (numerical) analysis of the thermodynamic stability of a given 
configuration $u(z)$ to the search for the normalizable eigenfunctions 
$\eta_{\Lambda}(z)$ in the finite interval of $\Lambda$, 
$\Lambda_{min}\leq \Lambda<0$.

Before considering Eq.~(\ref{eta}) with the general periodic function 
$u(z)$, let us establish the criterion for the thermodynamic stability 
of the particular homogeneous ({\em ferro} or {\em antiferro}) solution 
$u_c(z)=\pm\sqrt{-\lambda}$ of EL equation~(\ref{el}). Then Eq.~(\ref{eta}) 
reduces to the linear differential equation with constant coefficients, so that 
the normalizable eigenfunctions must have the form $\eta(z)\propto e^{ikz}$ 
with real values of the wave number $k$. The corresponding eigenvalues 
$\Lambda$ are given by 
\begin{equation}
\Lambda= k^{4} - k^{2} - 2\lambda \,\,\,,\,\,\, \lambda < 0.
\label{homog}
\end{equation}
It follows that the homogeneous configuration $u_c(z)=\pm\sqrt{-\lambda}$ 
is stable, i.e. that $\Lambda>0$ for any $k$, provided that $\lambda<-1/8$.
Note that the latter inequality is just the condition of {\em orbital} 
instability of the homogeneous solution. Namely, the linearization of 
the EL equation with respect to this solution leads to the linear equation 
\begin{equation}
\theta''''+\theta''- 2\lambda \theta=0,
\label{linel}
\end{equation}
which has normalizable solutions $\theta (z)$ only for $\lambda>-1/8$. Thus,
we see that in this simple case the thermodynamic stability excludes the 
orbital stability, and vice versa, and that two stabilities "meet" each other
in one point, $\lambda=-1/8$.

\subsection{General Floquet's procedure}

In order to apply the well-known Floquet's procedure \cite{yak} to 
Eq.~(\ref{eta}) with a general periodic function $u(z)$, we rewrite this 
equation in the matrix form
\begin{equation}
\frac{d{\bf \Theta}(z)}{dz}={\bf A}(z; \lambda, \Lambda){\bf \Theta}(z),
\label{etam}
\end{equation}
where ${\bf \Theta(z)} \equiv [\eta(z), \eta'(z), \eta''(z),\eta'''(z)]^T$,
and the matrix
$\mbox{{\bf A}}(z; \lambda, \Lambda)$ is given by
\begin{equation}
\mbox{{\bf A}}(z)=\left(\begin{array}{cccc}
 0  &  1  &  0  &  0 \\
 0  &  0  &  1  &  0 \\
0  &  0  &  0  &  1 \\
\Lambda-\lambda-3\,u(z)^{2} &  0  & -1  &  0  \end{array} \right).
\label{matrA}
\end{equation}
The system of linear equations~(\ref{etam}) has four linearly independent
solutions,  ${\bf \Theta}_i(z), i = 1, ...,4$. They form the fundamental 
matrix
\begin{equation}
 \mbox{{\bf F}}(z)=\left[{\bf \Theta}_{1}(z),{\bf
 \Theta}_{2}(z),{\bf \Theta}_{3}(z),{\bf \Theta}_{4}(z)\right],
\label{fm }
\end{equation}
which is obviously the solution of equation 
\begin{equation}
 \frac{d\mbox{{\bf F}}(z)}{dz}=\mbox{{\bf A}}(z;\lambda,\Lambda)\mbox{{\bf
 F}}(z).
\label{dfm}
\end{equation}
Without reducing generality we can always choose such initial conditions at 
$z = 0$ that $\mbox{{\bf F}}(0)=\mbox{{\bf I}}$, where $\mbox{{\bf I}}$ is the 
identity matrix.

The Floquet's theorem states that whenever the matrix $\mbox{{\bf
 A}}(z;\lambda,\Lambda)$ is a periodic function of variable $z$ with a 
 period $P$, the fundamental matrix has the form
\begin{equation}
 \mbox{{\bf F}}(z)=\mbox{{\bf G}}(z)\mbox{e}^{{\bf \Sigma}z},
\label{fg}
\end{equation} 
where $\mbox{{\bf G}}(z)$ is a matrix which varies periodically with $z$,
$\mbox{{\bf G}}(z+P)=\mbox{{\bf G}}(z)$, and ${\bf \Sigma}$ is a constant
matrix. Due to the periodicity of matrix $\mbox{{\bf G}}(x)$ and the 
initial condition $\mbox{{\bf G}}(0)=\mbox{{\bf I}}$, the matrix 
${\bf \Sigma}$ can be expressed in the form
\begin{equation}
 {\bf \Sigma}=\frac{1}{P}\ln \mbox{{\bf F}}(P).
\label{Sigma}
\end{equation}   
The matrix ${\bf F}(P)$ is called the monodromy matrix. The eigenvalues
of ${\bf \Sigma}$, $\sigma_i$ are Floquet's exponents and the eigenvalues of 
monodromy matrix $\mbox{{\bf F}}(P)$, $\rho_i$, are Floquet's multipliers. 
Put in other words, Floquet's theorem states that for each Floquet multiplier 
$\rho_i$ there exists a solution ${\bf \Theta}_i(z)$ of Eq.~(\ref{etam}) 
with the property 
\begin{equation}
{\bf \Theta}_i(z+P)=\rho{\bf \Theta}_i(z).
\label{floq}
\end{equation}
 
Floquet multipliers $\rho_i$ are in general complex numbers. For a 
normalizable solution ${\bf \Theta}_i(z)$ the Floquet multiplier $\rho_i$ 
lies on the unit circle in the complex $\rho$-plane, and the corresponding
Floquet exponent $\sigma_i$ is imaginary.

\subsection{Poincar\'{e}-Lyapunov theorem}
 
The problem~(\ref{etam}) has an additional important property. After the 
linear transformation ${\bf Z}(z) 
\equiv [Z_{1}(z), Z_{2}(z), Z_{3}(z),Z_{4}(z)]^T 
= {\bf T}{\bf \Theta}(z)$, defined by 
\begin{equation}
 Z_{1}=\eta(z)\,\,,\,\, Z_{2}=\eta''(z)\,\,,\,\, Z_{3}=2(\eta'(z)+\eta'''(z))
  \,\,,\,\, Z_{4}=2\eta'(z),
\label{tzeta}
\end{equation}\
i. e. by 
\begin{equation}
\mbox{{\bf T}}(z)=\left(\begin{array}{cccc}
 1  &  0 &  0  &  0 \\
 0  &  0  &  1  &  0 \\
0  &  2  &  0  &  2 \\
0&  2  & 0  &  0  \end{array} \right),
\label{matrT}
\end{equation}
Eq.~(\ref{etam}) is transformed into the equation 
\begin{equation}
\frac{d \mbox{{\bf Z}}(z)}{dz}=
\mbox{{\bf J H}}(z; \lambda, \Lambda)\mbox{{\bf Z}}(z),
\label{JH}
\end{equation}
with
\begin{equation}
 \mbox{{\bf  H}}(z)=\left(\begin{array}{cccc}
       -2(\lambda-\Lambda+3\,v_{0}(z)^{2})  &  0  &  0  &  0 \\
  0  &  2  &  0  &  0 \\
  0  &  0  &  0  & - \frac{1}{2} \\
  0  &  0  &- \frac{1}{2}  & \frac{1}{2}  \end{array} \right)\,\,,\,\,
\mbox{{\bf J}}=\left(\begin{array}{cccc}
  0 & 0 & -1 & 0 \\
  0 & 0 &  0 & -1 \\
  1 & 0 &  0 & 0 \\
  0 & 1 &  0 & 0 \end{array} \right).
\label{defJH}
\end{equation} 
The problem~(\ref{JH}) has the Hamiltonian form, characterized by the 
hermitean matrix $\mbox{{\bf H}}$ and the simplectic matrix $\mbox{{\bf J}}$ 
(i.e. $\mbox{{\bf J}}$ is antisymmetric and has the property  
$\mbox{{\bf J}}^2 = - \mbox{{\bf I}}$). The Poincar\'{e}-Lyapunov (PL) 
theorem~\cite{poin} for such problems states that the corresponding 
fundamental matrix, ${\bf \Phi}(z)$, satisfies the relation
\begin{equation}
{\bf \Phi}^{T}(z){\bf J}{\bf \Phi}(z)= {\bf \Phi}^{T}(0){\bf J}{\bf \Phi}(0).
\label{fijfi}
\end{equation} 
In other words, ${\bf \Phi}^{T}(z){\bf J}{\bf \Phi}(z)$ is the "integral of
motion" for the Hamiltonian problem~(\ref{JH}). This theorem
can be easily checked by differentiating the relation~(\ref{fijfi}) 
with respect to $z$, and taking into account Eq.~(\ref{JH}). 
Since ${\bf \Phi}(z) = \mbox{{\bf T}}\mbox{{\bf F}}(z)$, and the matrix 
$\mbox{{\bf T}}^T \mbox{{\bf J}} \mbox{{\bf T}} \equiv \mbox{{\bf J}}_{1}$
is also simplectic, it follows that 
\begin{equation}
{\bf F}^{T}(z){\bf J}_{1}{\bf F}(z) = {\bf F}^{T}(0){\bf J}_{1}{\bf F}(0) 
= {\bf J}_{1} ,
\label{fjf}
\end{equation}
i.e. the PL theorem holds for our original fundamental matrix 
$\mbox{{\bf F}}(z)$ as well.

From the relation~(\ref{fjf}) it follows that the matrices 
$\mbox{{\bf F}}^T(z)$ and $\mbox{{\bf F}}^{-1}(z)$ are similar 
[$\mbox{{\bf F}}^{T}(z) = \mbox{{\bf J}}_{1} \mbox{{\bf F}}^{-1}(z)
\mbox{{\bf J}}_{1}^{-1}$]. Thus, if $\rho_1\equiv \rho$ is the Floquet 
multiplier of  $\mbox{{\bf F}}(P)$, then $\rho^{-1}$ is also its Floquet 
multiplier. Furthermore, since in our example the matrix $\mbox{{\bf F}}(P)$ 
is real, it follows that $\rho^*$ and $\rho^{*-1}$ are Floquet multipliers 
as well. These simple relations link four Floquet multipliers of the 
problem~(\ref{etam}) for any periodic solution of EL equation~(\ref{el}) 
and for any value of parameter $\lambda$. The corresponding three possible 
types of distributions of Floquet multipliers in the complex $\rho$-plane are 
shown in Fig.~\ref{flmu}. Floquet multipliers are either complex (a) or real. 
In the latter case two pairs generally have different values and may be of the 
same (b) or opposite (c) signs. Fig.~\ref{flmu} does not include the 
situations with existing collective modes, i. e. when one or two pairs of 
solutions are normalizable, and the corresponding Floquet multipliers are on 
the unit circle.  

Four Floquet multipliers of the problem~(\ref{etam}) can be represented 
as roots of a polynomial function of fourth order. Since, due to the PL 
theorem, $\rho,\rho^{-1}, \rho^*$ and $\rho^{*-1}$ are all roots of such 
function, its general form is
\begin{equation}
P_{4}(\rho)=\rho^{4}+a(\lambda,\Lambda)\left(\rho^{3}+\rho\right)+
 b(\lambda,\Lambda)\rho^{2}+1,
\label{pol}
\end{equation} 
where $a(\lambda,\Lambda)$ and $b(\lambda,\Lambda)$ are, for a given periodic
function $u(z)$,  some smooth real functions of parameters $\lambda$ and 
$\Lambda$ from the matrix~(\ref{matrA}). 

\subsection{Scenarios of thermodynamic (in)stabilities}

Having recapitulated the Floquet theory for the Hamiltonian linear problem
(\ref{etam}), we address the problem of thermodynamical stability for a given
periodic configuration $u(z)$. We start by noting that Eqs.~(\ref{etam}), 
(\ref{matrA}) and (\ref{pol}) enable some general conclusions 
about the dependence of the positions of Floquet multipliers in the complex 
plane on the parameters $\lambda$ and $\Lambda$. At first, since
$\rho = 0$ cannot be the root of $P_{4}(\rho)$, it follows that by changing 
continuously $\lambda$ and $\Lambda$ one can come from the distributions (a) 
or (b) to the distribution (c) in Fig.~\ref{flmu} only by passing through the 
unit circle. Next, it is easy to determine the positions of Floquet multipliers 
in the limits $\Lambda \rightarrow - \infty$ and  $\Lambda \rightarrow \infty$,
since then we may neglect $\lambda$ and $u^2(z)$ with respect to $\Lambda$
in the polynomial matrix element of the matrix ${\bf A}$~(\ref{matrA})
[still keeping in mind that  $u(z)$ defines the period $P$ which enters into 
the definition~(\ref{Sigma})]. 

In the former limit $\Lambda \rightarrow - \infty$ the Floquet multipliers 
are given by
\begin{equation}
\rho_{n}= \mbox{e}^{k_{n}P}\,\,,\,\, 
k_{n}=\mid \Lambda \mid^{\frac{1}{4}}\mbox{e}^{i(2n+1)\frac{\pi}{4}}
\,\,,\,\, n=0,1,2,3\,
\label{minfty}
\end{equation}
i.e. the distribution from Fig.~\ref{flmu}(a) is realized. Note that for 
$\Lambda < \Lambda_{min}$ this distribution cannot pass to that from 
Fig.~\ref{flmu}(c), since in this range of values of $\Lambda$ the unit circle 
cannot be crossed because the problem (\ref{etam}) does not have normalizable 
solutions. 

In the limit $\Lambda \rightarrow \infty$ the Floquet multipliers are given by
\begin{equation}
\rho_{1} = \rho_{2}^{-1} = \mbox{e}^{\Lambda^{\frac{1}{4}}P}\,\, \,\,,\,\,\,\,
\rho_{3} = \rho_{4}^{-1} =\mbox{e}^{i\Lambda^{\frac{1}{4}}P}\,\,.
\label{infty}
\end{equation}
As is seen in Fig.~\ref{lain}, one then has one pair of normalizable and one 
pair of non-normalizable solutions. 

From the other side, at $\Lambda = 0$ one particular Floquet multiplier
has the value $\rho = 1$, and corresponds to the already mentioned Goldstone 
mode. It has to be at least doubly degenerate, since otherwise the 
remaining three multipliers could not have symmetric positions required
by the PL theorem. Possible distributions of Floquet multipliers for  
$\Lambda = 0$ are shown in Fig.~\ref{lam0}. Aside from the possibility that the
degeneracy of the Goldstone mode is complete and all four multipliers are 
at $\rho = 1$ (a), one may have the remaining two multipliers either
on the real axis (b,c), or on the unit circle (d).

Taking into account the above conclusions, we are now able to list 
possible scenarios of thermodynamic (in)stabilities for the periodic
solutions of Eqs.~(\ref{el},\ref{condA}).
\renewcommand{\labelenumi}{\roman{enumi})}
\begin{enumerate}

\item The solution $u(z)$ is unstable for a given value of $\lambda$ if by 
increasing $\Lambda$ from $\Lambda_{min} = \lambda-1/4$ the Floquet multipliers
from the distribution (a) or (b) of Fig.~\ref{flmu} move in such a way to 
come to the unit circle for some value of $\Lambda$ in the interval 
$(\lambda-1/4, 0)$.

\item If the solution $u(z)$ is thermodynamically stable, the 
Floquet multipliers for $\lambda-1/4 < \Lambda < 0$ are defined either by one
complex number not lying on the unit circle (Fig.~\ref{flmu}a), or by two real 
numbers of the same sign  ($r_1, r_2$) and their reciprocals 
(Fig.~\ref{flmu}b). The 
latter case has to be realized as $\Lambda \rightarrow 0$ from below, since 
only distributions (b) and (c) from Fig.~\ref{lam0} represent the Goldstone 
mode for a thermodynamically stable configuration. Thus, at some negative 
value of $\Lambda$ the distribution from  Fig.~\ref{flmu}a has to reduce to 
the double degenerate Floquet multiplier at the real axis ($r_1 = r_2$), 
which then evolves into the distribution from Fig.~\ref{flmu}b.

\item For special value(s) of control parameter 
($\lambda = \lambda_c$) the thermodynamic instability of $u(z)$ proceeds 
in a particular way, realized when all four complex Floquet multipliers 
approach together the point $\rho_0 = 1$  as $\Lambda$ tends to zero from 
below. The Goldstone mode is then completely degenerate (Fig.~\ref{lam0}a). 
Putting in another way, such instability occurs when the 
points $r$ and $r^{-1}$ in Fig.\ref{lam0}b tend towards $\rho_0 = 1$ as 
$\lambda \rightarrow \lambda_c$. Note that the distribution of Floquet 
multipliers from Fig.~\ref{lam0}c means that the instability, i.e. the 
crossing of Floquet multipliers with the unit circle, takes place at some 
negative value of $\Lambda$. Also, the distribution from Fig.~\ref{lam0}d 
signifies that the remaining non-Goldstone mode is unstable in a finite 
interval of values of $\Lambda$, starting at some negative value of $\Lambda$. 
\end{enumerate}

\section{Bloch theory}

In order to determine normalizable solutions of Eq.~(\ref{etam}) with 
$\Lambda \geq 0$, i.e. the collective modes for given periodic
configuration $u(z)$ with the period $P = 2\pi/Q$, we profit from the freedom 
in choosing boundary conditions for the solutions $\eta(z)$, and specify
periodic (Born - von Karman) ones. By this we chose the Bloch representation, 
\begin{equation}
\eta_{k}(z)=e^{ikz}\Psi_{k}(z),\,\,\,\,\,\,\,\,
\Psi_{k}\left(z+\frac{2\pi}{Q}\right)=\Psi_{k}(z),
\label{bloch}
\end{equation}
where $k$ is the Bloch wave number limited to the I Brillouin zone 
($-Q/2\leq k \leq Q/2$). The differential equation for the periodic 
function $\Psi_{k}(z)$ reads
\begin{eqnarray}
& & \frac{d^{4}\Psi_{k}(z)}{dz^{4}}+4ik
\frac{d^{3}\Psi_{k}(z)}{dz^{3}}+(1-6k^{2})\frac{d^{2}\Psi_{k}
 (z)}{dz^{2}}+2ik(1-2k^{2})\frac{d\Psi_{k}(z)}{dz}+\nonumber \\
& &+ \left[k^{4}-k^{2}+\lambda+3u(z)^{2}\right]\Psi_{k}(z)=
\Lambda(k)\Psi_{k}(z),
\label{eqpsi}
\end{eqnarray}
and the normalizability condition is
\begin{equation}
\frac{Q}{2\pi}\int_{0}^{\frac{2\pi}{Q}}\Psi_{k}^{*}(z)\Psi_{k}(z)dz=1.
\label{normpsi}
\end{equation}
The dependence $\Lambda(k)$, i. e. the spectrum of eigenvalue 
problem~(\ref{eta}), follows from Eqs.~(\ref{eqpsi}, \ref{normpsi}). Since $k$ 
is quasi-continuous in the limit $L \rightarrow \infty$, this spectrum is for 
a stable configuration $u(z)$ composed of non-negative bands. The corresponding 
Bloch functions $\eta_{n,k}(z)$, where $n$ enumerates bands, represent a 
complete orthonormal set of functions for the problem~(\ref{eta}).

From expressions~(\ref{floq}) and~(\ref{bloch}) it follows that the
Floquet multiplier for the Bloch function $\eta_{k}(z)$ is given by  
$\rho=\mbox{e}^{ikP}$. The polynomial function~(\ref{pol}) then has the form 
\begin{equation}
P_{4}(\rho)=\left(\rho-\mbox{e}^{ikP}\right)
\left(\rho-\mbox{e}^{-ikP}\right)
\left[\rho^{2}+c_{2}(\lambda,\Lambda)\rho + 1\right]\,,  
\label{polbloch}
\end{equation}
where $c_{2}(\lambda,\Lambda)$ is some coefficient. Comparing two
representations for $P_{4}(\rho)$ we conclude that the coefficients from 
the expressions~(\ref{pol}) and~(\ref{polbloch}) are linked by relations
\begin{equation}
a(\lambda,\Lambda) = c_{2}(\lambda,\Lambda) - 2 \cos(kP)\,,\,\,\,\,\,\,\,\,\,
b(\lambda,\Lambda) = 2 - 2 c_{2}(\lambda,\Lambda) \cos(kP)\,,
\label{coeff}
\end{equation}
i. e. that the eigenvalue $\Lambda$ depends on the wave number $k$ only 
through the function $\cos(kP)$. This in particular means that for each
band $\Lambda(k)$ we have $\Lambda(-k)=\Lambda(k)$. This is consistent with 
the symmetry of Eq.~(\ref{eqpsi}). Furthermore
$\Lambda(k+\frac{2\pi}{P})=\Lambda(k)$, in accordance with the reduction
of wave numbers in~(\ref{bloch}) to the I Brillouin zone.

The representation~(\ref{bloch}) is particularly convenient for the analytical 
discussion of long wavelength limit $k \rightarrow 0$ for the Goldstone 
mode for which $\Lambda(k=0)=0$ and $\Psi_{k=0}(z)=u'(z)$. To this end we 
insert the Taylor expansions for small $k$,
\begin{equation}
\Psi_{k}(z)=u'(z)+k\Psi_{1}(z)+k^{2}\Psi_{2}(z)+...\,\,\,,\,\,\,
\Lambda(k)=k^{2}\Lambda_{2}+k^{4}\Lambda_{4}+... ,
\label{ex1}
\end{equation}
into Eq.(~\ref{eqpsi}). The requirement that the coefficients in front of 
leading powers, $k$ and $k^{2}$, vanish then leads to the equations  
\begin{equation}
\left(\tilde{\cal{D}}^{2}+\lambda-\frac{1}{4}+3u^{2}\right)\Psi_{1}=
                  -2i\left(2u''''+u''\right)\,\,,
\label{ex2}
\end{equation}
and
\begin{equation}
\left(\tilde{\cal{D}}^{2}+\lambda-\frac{1}{4}+3u^{2}\right)\Psi_{2}=
                  (\Lambda_{2}+1)u'+6u'''-2i(2\Psi_{1}'''+\Psi_{1}'),
\label{ex3}
\end{equation}
with the operator $\tilde{\cal{D}}$ given by Eq.~(\ref{etah}).
After multiplying 
equation~(\ref{ex3}) by $v'$, integrating with respect to $z$, using the fact  
that $u'$ is the Goldstone mode, and inserting $2u''''+u''$ from
Eq.~(\ref{ex2}), we get the expression for the coefficient $\Lambda_{2}$,
\begin{equation}
\Lambda_{2}=-1+6\frac{\left<u''^{2}\right>}{\left<u'^{2}\right>}-           
  \frac{\left<\Psi_{1}^{*}\left(\tilde{\cal{D}}^{2}+\lambda-\frac{1}{4}+3u^{2}
              \right)\Psi_{1}\right>}{\left<u'^{2}\right>},
\label{ex4}
\end{equation}
where $\left<...\right>$ stands for the spatial integration, like in 
Eq.~(\ref{norm}). However, the general thermodynamic condition~(\ref{condA})
for the functional ~(\ref{fe1}) reads~\cite{dbprl}
\begin{equation}
\frac{\left<u''^{2}\right>}{\left<u'^{2}\right>} = \frac{1}{2}\,\,,
\label{ex5}
\end{equation}
so that Eq.~(\ref{ex4}) can be written in a more transparent way,
\begin{equation}
\Lambda_{2}=2-F_{2}\left[u(z)\right]\,\,.
\label{ex6}
\end{equation} 
Here we introduce the functional
\begin{equation}
F_{2}\left[u(z)\right]=
\frac{\left<\Psi_{1}^{*}\left(\tilde{\cal{D}}^{2}+\lambda-\frac{1}{4}+3u^{2}
              \right)\Psi_{1}\right>}{\left<u'^{2}\right>}.
\label{ex7}
\end{equation}
Since the operator figuring in this equation just defines the eigenvalue
problem~(\ref{eta},\ref{etah}), it is clear that the functional 
$F_{2}\left[u(z)\right]$ is positive definite for any thermodynamically stable 
configuration $u(z)$. This has two consequences.

Firstly, the common upper limit of the velocity of Goldstone mode, 
$v_{G}=\sqrt{\Lambda_{2}}$, for all thermodynamically stable periodic states 
is $v_{G,M}=\sqrt{2}$. The velocity $v_{G}$ for a given (meta) stable state 
has the maximum value $v_{G,max}\leq v_{G,M}$ when the functional~(\ref{ex7}) 
attains 
its minimum. Like the functional~(\ref{fe2}), the functional $F_{2}[u(z)]$ 
depends only on the parameter $\lambda$. Thus, taken a given solution $u(z)$,
we can find the function $\Psi_1$ by solving the inhomogeneous linear 
differential equation~(\ref{ex2}), and then determine, by calculating 
$F_{2}[u(z)]$, the velocity  $v_{G}$ as a function of $\lambda$. In other 
words, we have a direct method for the calculation of the velocity of  
Goldstone mode, not related to the above Floquet-Bloch procedure (but derived 
from it). It can be used as an independent check of numerical results for 
the spectrum $\Lambda(k)$ which follow from Eq.~(\ref{eqpsi}).

Secondly, the functional~(\ref{ex7}) attains its minimal value ($F_2 = 0$)  
{\em if and only if} the function $\Psi_{1}(z)$ vanishes. As is seen from
Eq.~(\ref{ex2}), this is possible only when the solution $u(z)$ satisfies the 
equation
\begin{equation}
 2u''''+u''=0.
\label{ex8}
\end{equation} 
i.e. when $u(z)\propto \sin{(z/\sqrt{2})}$. The only solution from
the phase diagram in Fig.~\ref{phdi} with this property is the {\em almost} 
sinusoidal incommensurate state, denoted by $s^2$. Since in the limit 
$\lambda \rightarrow \lambda_{id} = 1/4$ it reduces strictly to the above 
simple sinusoidal dependence on $z$
(with the amplitude tending to zero), we conclude that just at the second 
order phase transition from the incommensurate to the disordered state the 
velocity of Goldstone mode of incommensurate state attains the maximum 
value $v_{G,M} = \sqrt{2}$.  We note that other periodic (and metastable) 
states $u(z)$ from Fig.~\ref{phdi} cannot even approximately satisfy 
Eq.~(\ref{ex8}). From the other side, due to the deviations from the sinusoidal 
form of a given solution, the functional~(\ref{ex7}) can attain
the value $F_{2}=2$, in which case the velocity of Goldstone mode vanishes.
As will be seen from the numerical results in the next section, this is indeed
the case for all periodic solutions (including the almost sinusoidal 
configuration $s^2$) at the edges of their local thermodynamic stabilities.  

Let us conclude this general discussion with the remark on the class of 
periodic solutions $u(z)$ which in addition have the property 
$u(z+P/2)= - u(z)$. Since only $u^{2}(z)$, which then has the period 
$P/2$ (and not $P$), enters into the problem~(\ref{etam}), the corresponding 
spectrum $\Lambda(k)$ and the Floquet multipliers can be calculated with 
respect to the former period. The I Brillouin zone is then doubled 
($-Q\leq k \leq Q$), and the number of branches of collective modes is 
halved. In particular, within this choice the value of Floquet 
multiplier for the Goldstone mode $u'(z)$, defined by Eq.~(\ref{floq}), 
is $-1$ and not $1$. The approaching of $\Lambda = 0$ from below for the 
stable solution $u(z)$ then proceeds like in the point ({\em ii}) of Subsection
IV.A, but with one pair of Floquet multipliers tending towards the point 
$\rho_{0} = -1$, and the other pair placed at the negative real semiaxis 
(Fig.~\ref{lam0}c). Correspondingly, the Bloch representation of  
Goldstone solution of Eq.~(\ref{etam}) is $\eta = \mbox{e}^{\pm iQz} \Psi(z)$ 
with $\Psi(z) = \mbox{e}^{\mp iQz}u'(z)$, i.e. the Goldstone mode is 
placed at the border of the doubled Brillouin zone. However, the
propagation of collective modes takes place in the periodic structure
determined by the configuration $u(z)$ (and the period $P$). Thus the 
above doubled Brillouin zone has to be folded once to get the physical one,
$-Q/2\leq k \leq Q/2$. In other words, the wave numbers $k = Q$ and $k = 0$ 
coincide, so that the Goldstone mode is realized as the long-wavelength one 
for such solutions as well. Furthermore, we note that after this folding the 
above Taylor expansion~(\ref{ex1}) and subsequent conclusions on the velocity 
of Goldstone mode [Eqs.~(\ref{ex6}-\ref{ex8})] follow in the same way for 
states with the property $u(z+P/2)= - u(z)$ as well.

\section{Collective modes for systems of class II}

In order to derive collective modes for configurations participating in the
phase diagram from Fig.~\ref{phdi}, 
we extend the numerical method developed in Refs.~\cite{dbpra,dbpre} to the 
calculation of eigenvalues and Bloch solutions~(\ref{bloch}) of 
the linear problem~(\ref{eta}, \ref{etam}). In the further discussion we shall 
mostly use the notation $\Omega(k)\equiv \sqrt{\Lambda(k)}$, where $\Omega(k)$ 
has the meaning of the frequency of collective mode. Note that the energy 
scale for $\Omega(k)$ [as well as that for averaged free energies in
Fig.~\ref{phdi}] is defined by the last expression in Eq.~(\ref{red}).
 
The periodic solutions from the phase diagram were determined by 
solving a system of algebraic equations for coefficients of their Fourier 
expansions. These Fourier sums were truncated at finite degrees, high enough 
to ensure a sufficient precision for $u(z)$, as well as for corresponding 
wave number $Q$, averaged free energy, etc. The limitation of this method 
comes from the increase of number of non-negligible Fourier coefficients 
as the period $2\pi/Q$ increases, and the corresponding solutions $u(z)$ 
contain more and more elementary sinusoidal and uniform segments. 

Representing the function  $u^2(z)$ in Eq.~(\ref{eqpsi}) by corresponding 
truncated Fourier series, and writing the function $\Psi_k(z)$ in the same 
manner, 
\begin{equation} 
\Psi_{k}(z) = a_{0}+\sqrt{2}\sum_{n=1}^{N}\left[a_{n}\cos (nQz) + 
b_{n}\sin(nQz)\right],
\label{fourpsi}
\end{equation}
we come to the homogeneous linear algebraic system for the coefficients 
$a_{0},a_{1},..,a_{N}$ and $b_{1},b_{2},...,b_{N}$. In order to calculate 
collective modes $\Lambda(k)$, it remains to diagonalize the corresponding 
($2N + 1$) - dimensional matrix. This matrix is generally complex and 
hermitean. Again, one has to keep a sufficient number of Fourier components in 
the expansion~(\ref{fourpsi}) to get a reliable result for at least two lowest 
branches in the spectrum $\Lambda(k)$. In actual calculations the truncation 
at a given number of coefficients $N$ is taken as acceptable if for a given 
branch $\Lambda(k)$ one fulfills to a certain degree of approximation the 
equality $\Lambda(k=0) = \Lambda(k=Q)$ [i.e. the equality 
$\Lambda(k=0) = \Lambda(k=2Q)$ for the solutions with the property
$u(z+P/2)=-u(z)$]. 
 
\subsection{Collective modes of states $u_s(z)$, $u_c(z)$ and $u_d(z)$}

In Fig.~\ref{como} we present the spectrum of collective modes for the 
incommensurate almost sinusoidal state $u_s(z)$, denoted by $s^2$ in 
Fig.~\ref{phdi}, choosing few 
characteristic values of the parameter $\lambda$. As announced above, we use 
here the reduced Brillouin zone, $-Q/2 < k < Q/2$, for all values of $\lambda$, 
except for those for which the periodic modulation is absent 
($\lambda = 0.3 > \lambda_{id}$ in Fig.~\ref{como}c). At the very second order 
transition from the incommensurate state  to the disordered state $u_d(z) = 0$
($\lambda = \lambda_{id} = 1/4$) we present the spectrum in both, reduced and 
extended, zone schemes (Fig.~\ref{como}c). Note that due to the additional 
symmetry of $s^2$ state, $u_s(z+\pi/Q) = - u_s(z)$, the subsequent branches in 
Figs.~\ref{como}a-b are not separated by gaps at the zone edges $k=\pm Q/2$. 

At first, we see that
the lowest branch has the property of Goldstone mode [$\Omega(k) \sim k$ 
for $k \rightarrow 0$] in the whole range of stability of the configuration
$s^2$. For $\lambda$ well below the critical value $\lambda_{id}$
($\lambda= - 0.1$ in Fig.~\ref{como}a) the subsequent pairs of 
branches defined in such way are separated by gaps at $k = 0$. 
In other words, the general property obtained before in the limit 
$\Lambda = \Omega^2 \rightarrow \infty$ by which only one pair of Floquet 
multipliers is on the unit circle (Fig.~\ref{lain}), is here realized for all 
values of $\Lambda$. However, as $\lambda$ increases the gap between two lowest 
pairs of branches decreases, and finally disappears for $\lambda \approx 0.05$, 
as is seen in Fig.~\ref{como}a. Then one has an overlap of branches in a 
finite range of values of $\Omega$, i.e. all four Floquet multipliers are on 
the unit circle. This overlap increases, and the minimum of higher branch 
tends towards $0$, as $\lambda$ approaches the critical value $\lambda_{id}$ 
($\lambda = 0.249$ 
in Fig.~\ref{como}b). At $\lambda = \lambda_{id}$ this minimum has 
the value $\Omega = 0$, while the slope $d\Omega(k)/dk$ has a finite value 
which coincides with that of already existing Goldstone branch 
(Fig.~\ref{como}c). In other words, just at the second order phase transitions 
one has two acoustic modes, which, although with same phase velocities 
$ v \equiv \lim_{k \rightarrow 0} \frac{d\Omega(k)}{dk}$, have 
different dispersions at finite values of $k$. For $\lambda > \lambda_{id}$ 
these two branches combine into a single mode which has minima at $k = \pm Q$ 
with a finite value $\Omega(Q)$, and a maximum at $k = 0$ (Fig.~\ref{como}c), 
as it follows directly from the quadratic part of Landau expansion~(\ref{fe2}).

The dependence of the phase velocity of Goldstone mode, $v_G$, on the 
parameter $\lambda$ is shown in Fig.~\ref{vgsi}. It is finite at 
$\lambda = \lambda_{id}$, decreases as the amplitude of the incommensurate 
state increases, and vanishes at the metastability edge for the $s^2$ state,
$\lambda = - 1.835$. This dependence is in accordance with the analytic 
results~(\ref{ex4}-\ref{ex8}) on the asymptotic behavior of the Goldstone mode. 

The spectrum of collective modes for the commensurate state 
$u_c(z) = \pm \sqrt{-\lambda}$ [i.e. $\tilde u_c(\tilde z) = \pm \sqrt{-a/b}$ 
in the original notation of Eq.~(\ref{fe1})] follows from Eq.~(\ref{homog}). 
This state is thermodynamically stable in the range 
$a < 0$ for $c > 0$ and $a < -c^2/(8d)$ for $c < 0$, which comprises positive 
values of $c$, excluded from the analysis after the transformation~(\ref{red}).
In order to cover the whole range of stability of $\tilde u_c(\tilde z)$, we 
rewrite Eq.~(\ref{homog}) in the original notation,
\begin{equation}
\tilde \Omega^2 = d \tilde k^4 + c \tilde k^2 - 2 a = d \left(\tilde k^2
+ \frac{c}{2d}\right)^2 - \frac{c^2}{4d} - 2a 
\label{homor}
\end{equation} 
[with $\tilde k \equiv \sqrt{-c/d}\, k, \,\,\tilde \Omega \equiv \sqrt{c^2/d} 
\,\Omega$ in the range $c < 0$]. The second equality in the
expression~(\ref{homor}) shows that for  $c < 0$ the dispersion curve
has minima at $\tilde k = \pm \sqrt{-c/(2d)}$ \,\,\,(i.e. at $k = 
\pm 1/\sqrt{2}$), with the gap $\tilde \Omega(\tilde k)$
equal to $\sqrt{-2a-c^2/(4d)}$ [i.e. $\sqrt{2(-\lambda - 1/8)}$ in the reduced 
scale $\Omega$]. As for the range $c > 0$, it follows from the first equality 
in Eq.~(\ref{homog}) that the collective mode has the minimum at $\tilde k = 0$,
with the gap $\tilde \Omega(0) = \sqrt{-2a}$. The gap vanishes at $a = 0$, i.e. 
at the line of second order transition from the commensurate state 
$\tilde u_{c}(\tilde z)$ to the disordered state $\tilde u_{d}(\tilde z) = 0$. 

The commensurate solutions $\tilde u_c(\tilde z) = \pm \sqrt{-a/b}$, which 
here represent the uniform or dimerized ordering for the Landau 
expansions~(\ref{fe1}) around the center or the border of the original 
Brillouin zone respectively, have the same symmetry properties as the solution 
for the disordered state, $\tilde u_d(\tilde z) = 0$. The only 
collective excitations with finite activated frequencies are fluctuations of 
the amplitude $\tilde u(\tilde z)$ with the above dispersion
relation~(\ref{homor}). Note that the mode of Goldstone (acoustic) type is 
absent. Since the solutions $\tilde u_c(\tilde z)$ possess, as constants, a 
trivial translational degeneracy, we prefer to associate this absence of  
acoustic branch with its reduction to the trivial dependence 
$\tilde \Omega (\tilde k) = 0$. The purpose of this interpretation will become 
clear in the next subsection. 

Finally, as it follows directly from the expression~(\ref{fe1}), the disordered 
state $\tilde u_d (\tilde z) = 0$ which is stable in the range 
$a > 0, c > - \sqrt{4ad}$, has a branch of collective excitations with the 
minimum at $\tilde k = 0$ for $c > 0$, and with two minima at  
$\tilde k = \pm \sqrt{-c/2d}$ for $c < 0$. The respective gaps at these minima
are equal to $\sqrt{a}$ (for $c > 0$), and to $\sqrt{a - c^2/4d}$ (for $c < 0$). 

\subsection{Collective modes of periodic metastable states}

\begin{figure}
\caption{The phase diagram for the model (\ref{fe2}). The inset shows
 the whole range of (meta)stability of the {\em almost} sinusoidal
 configuration $s^{2}$ and indicates the portion of the phase diagram in which
 other metastable non-sinusoidal configurations coexist. This portion is
 shown in the larger part of the figure.}
\label{phdi}
\bigskip
\bigskip
\bigskip
\bigskip
\bigskip
\centerline{\psfig{file=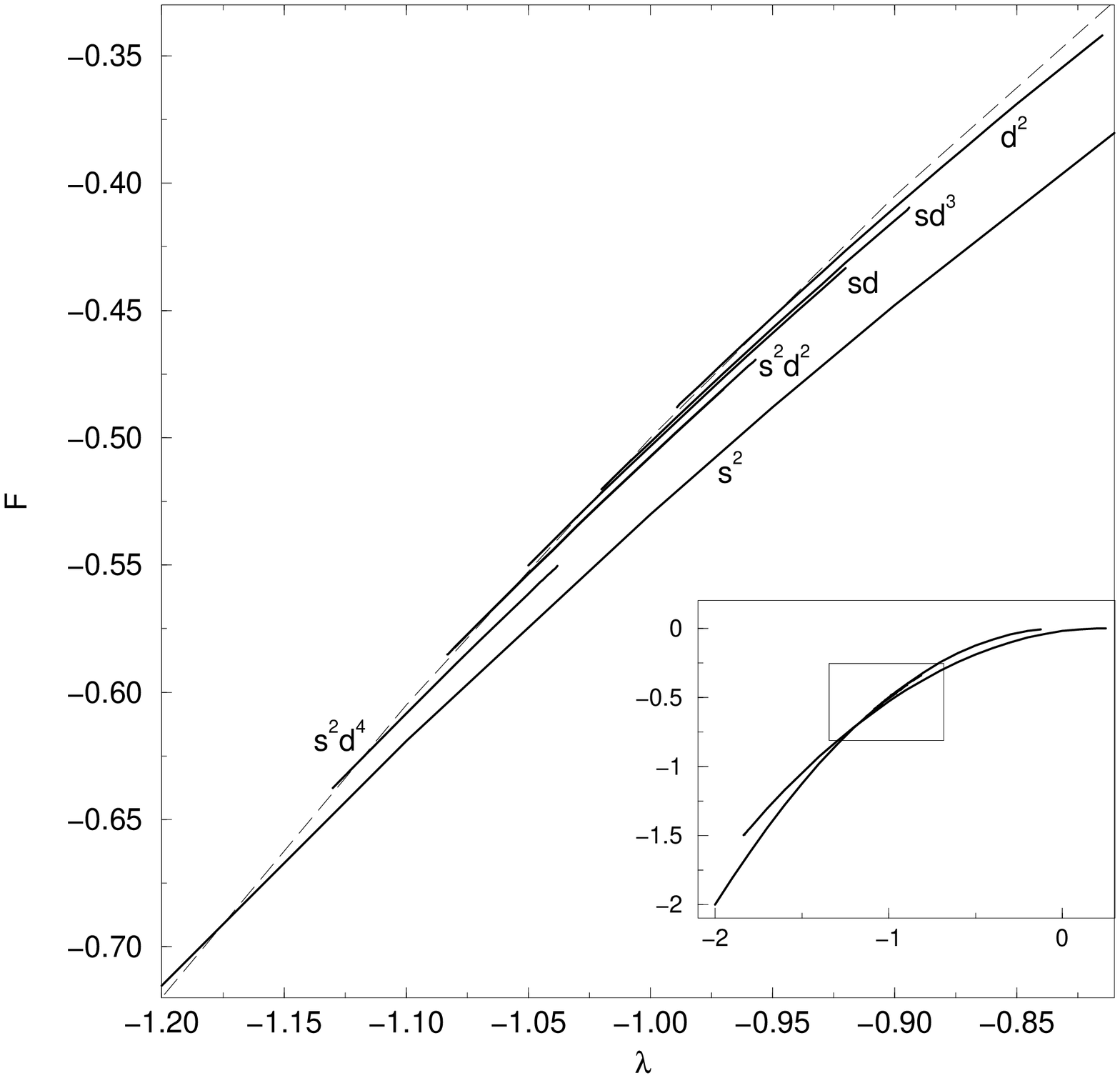,height=10cm,width=10cm,silent=}}
\end{figure}

\begin{figure}
\caption{Possible distributions of Floquet multipliers in the complex plane.
 The distributions (a), (b) and (c) correspond to the non-normalizable
 solutions of Eq.(\ref{etam}). None of these distributions can
 smoothly transform to other two without intersecting the unit circle.  }
\label{flmu}
\bigskip
\centerline{\psfig{file=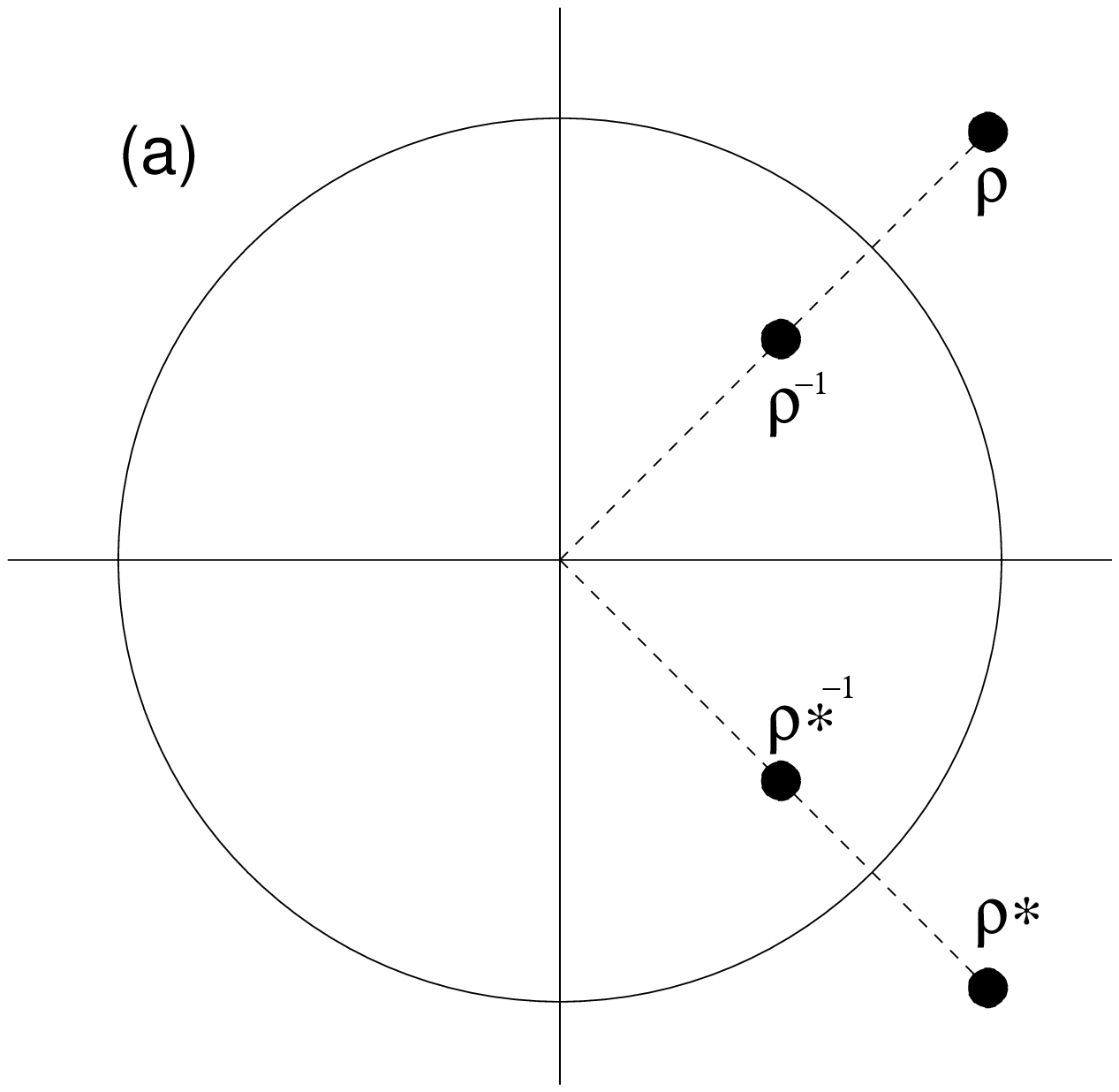,height=4.1cm,width=4.1cm,silent=}\hspace*{1cm}
\psfig{file=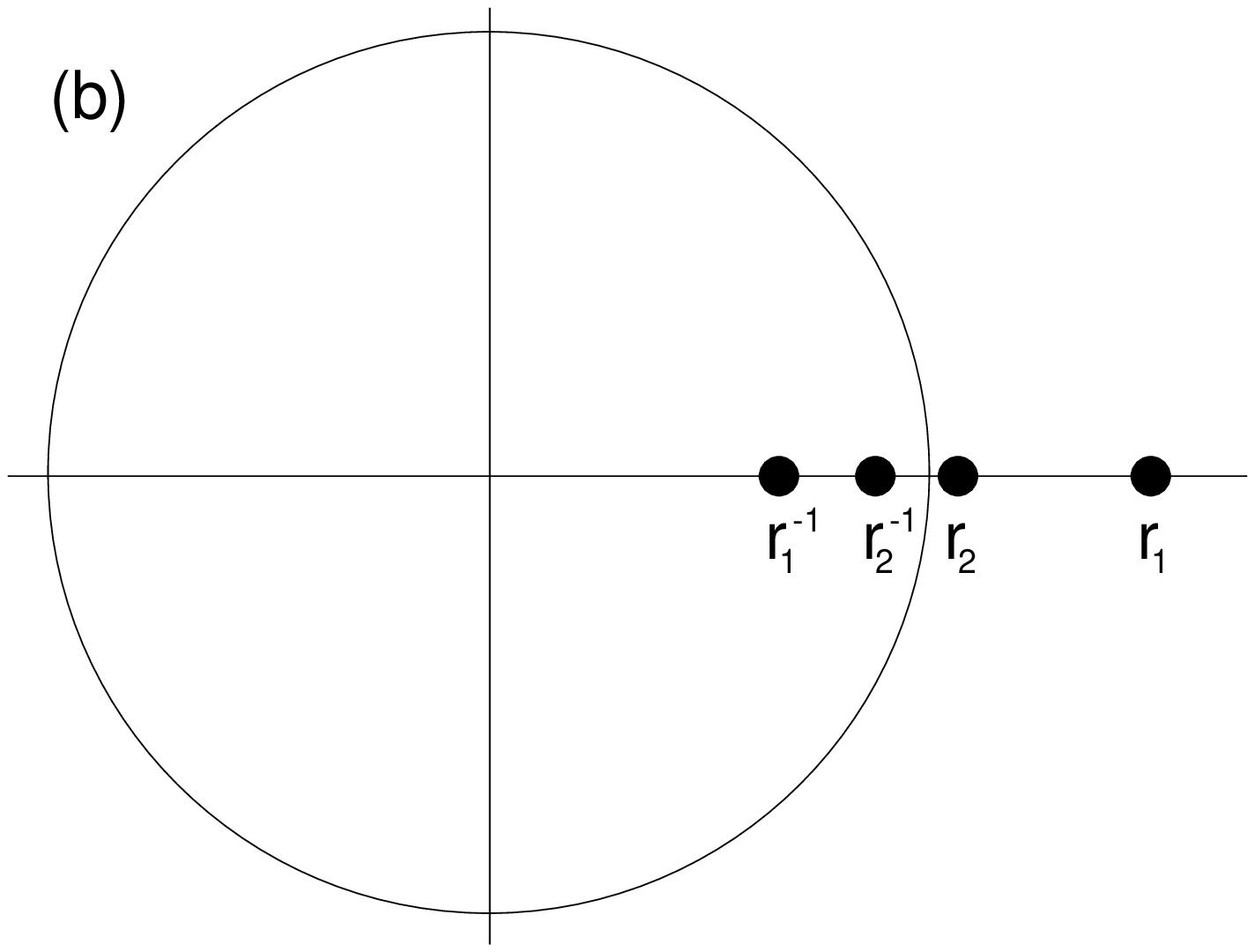,height=3.8cm,width=4.8cm,silent=} \hspace*{1cm}
\psfig{file=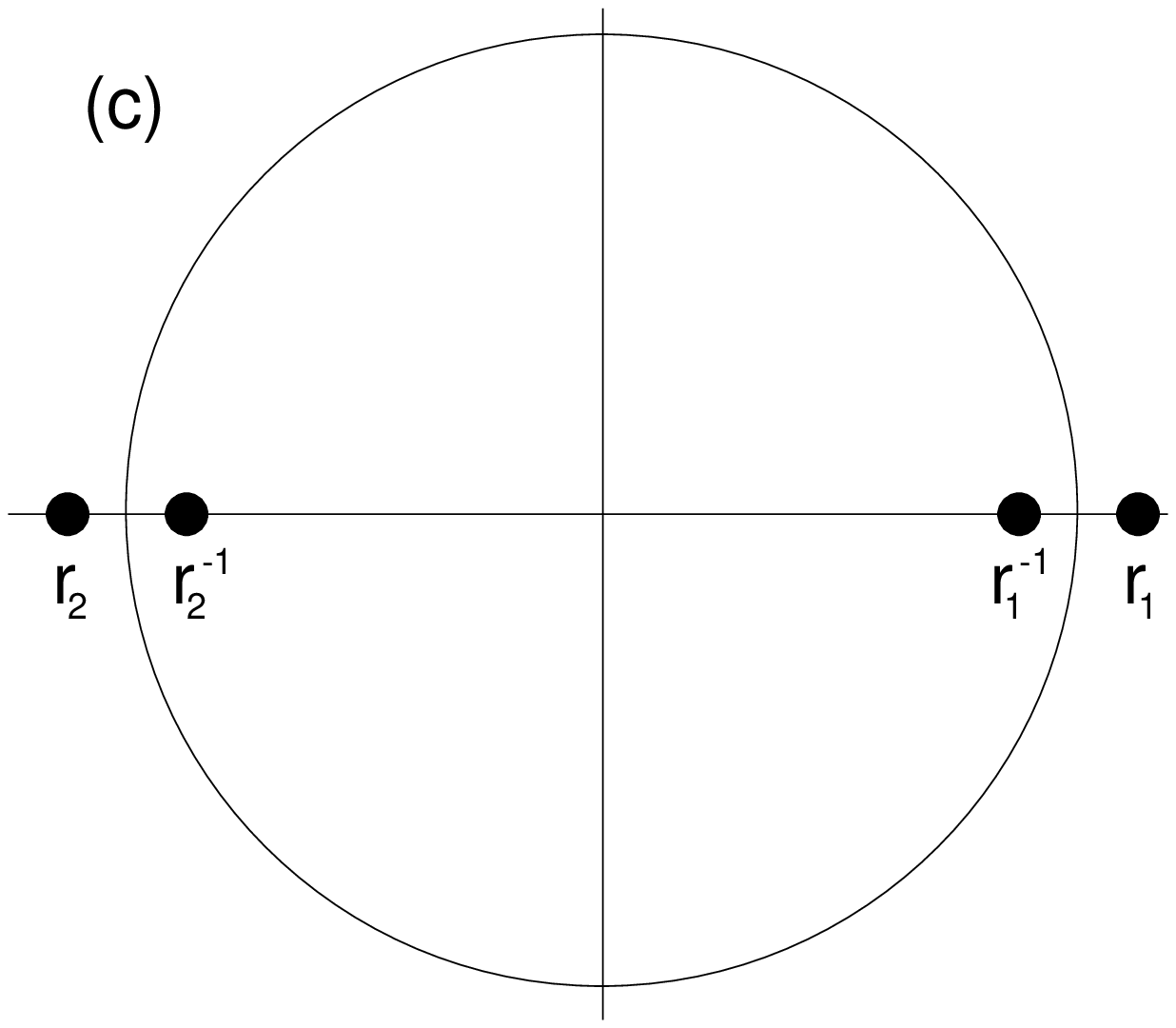,height=3.9cm,width=4.1cm,silent=}}
\end{figure}

\begin{figure}
\caption{The distribution of Floquet multipliers corresponding to the existence
 of just one complex normalizable solution (a collective mode) of 
 Eq.(\ref{etam}).}
\label{lain}
\centerline{\psfig{file=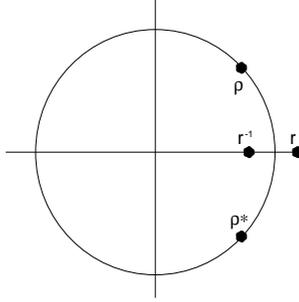,height=4cm,width=4cm,silent=}}
\end{figure}

\begin{figure}
\caption{The completely degenerate Goldstone mode (a) which originates by
 merging two real multipliers shown in (b). The situation shown in (c)
 corresponds to approaching the instability for some negative $\Lambda$, which
 then evolves by moving two multipliers (corresponding to a non-Goldstone 
 mode) along the unit circle.  }
\label{lam0}
\bigskip
\centerline{\psfig{file=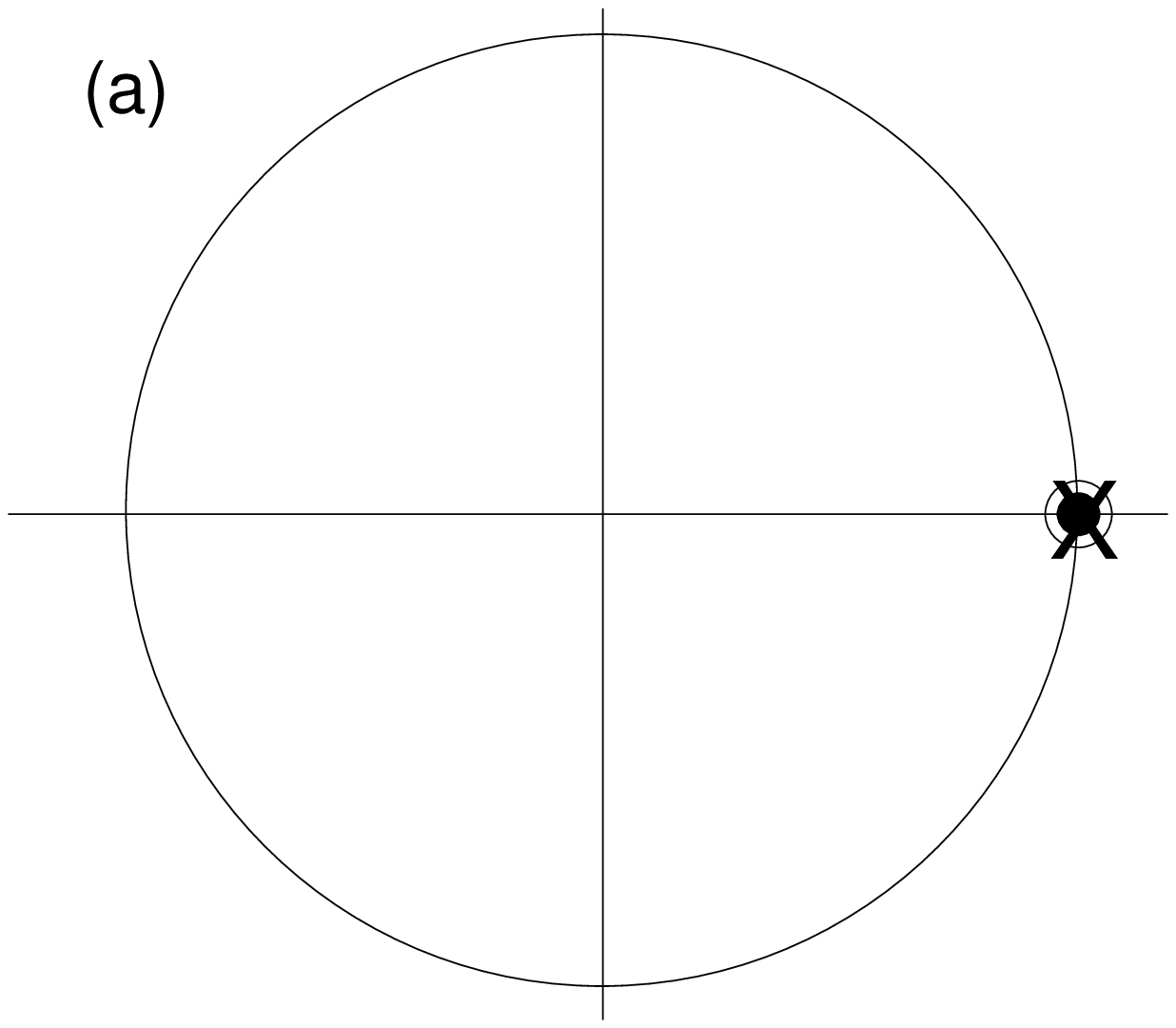,height=3.6cm,width=3.8cm,silent=}\hspace*{4mm}
\psfig{file=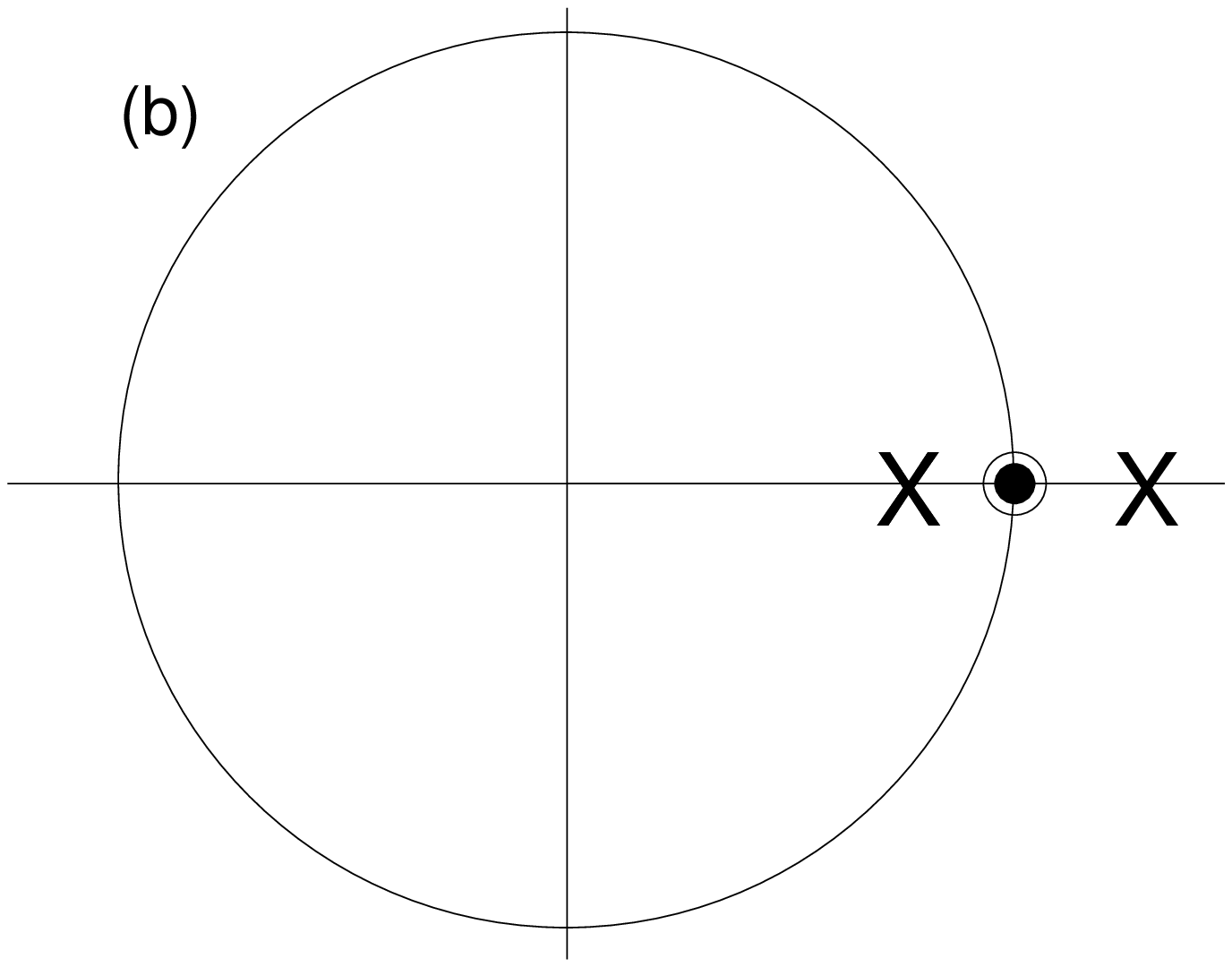,height=3.6cm,width=4.4cm,silent=}\hspace*{4mm}
\psfig{file=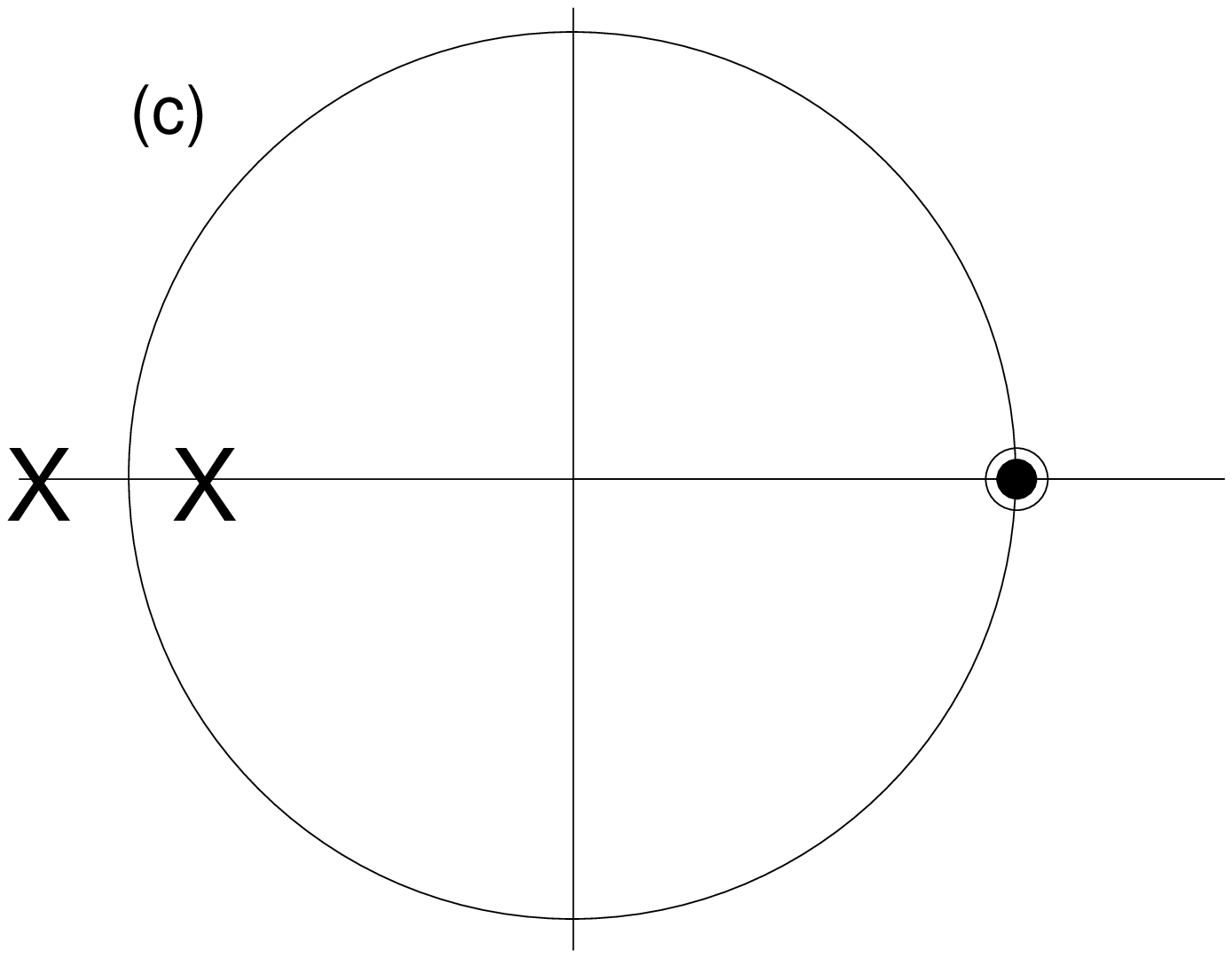,height=3.6cm,width=4.4cm,silent=}\hspace*{4mm}
\psfig{file=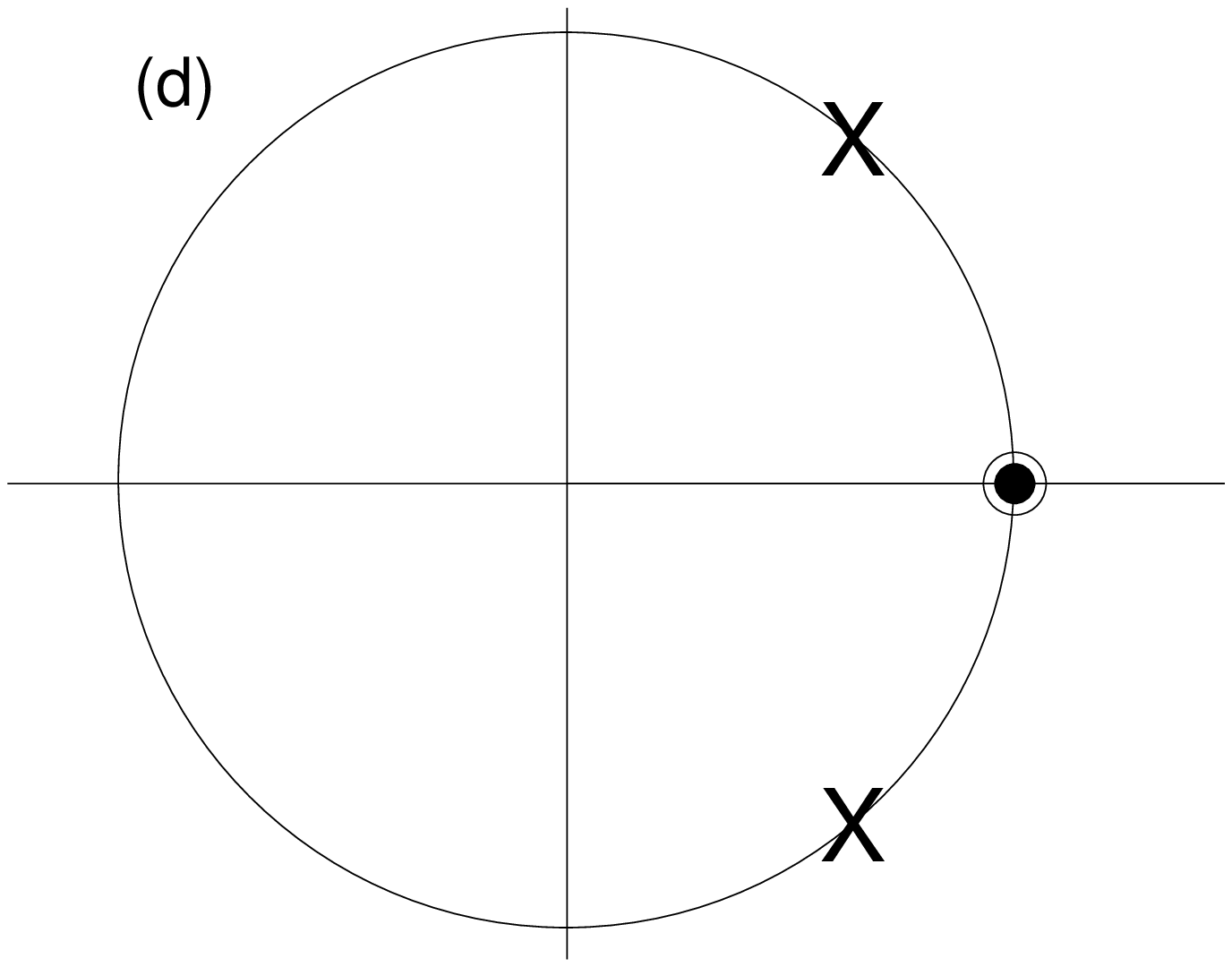,height=3.6cm,width=4.2cm,silent=}}
\end{figure}

\begin{figure}
\caption{ The dispersion curves for the almost sinusoidal configuration for
 different values of the control parameter $\lambda$. The curve (a) shows the
 situation when two branches are separated ($\lambda< 0.05$). By
 increasing $\lambda$ they begin to overlap for $\lambda\approx 0.05$ (b),
 until $\lambda$ reaches
 the critical value $\lambda=0.25$. The curve (c) corresponds to
 $\lambda=0.249$. The finite slope of curves at $k=0$ coincides with that of
 the Goldstone mode. For $\lambda>0.25$ (d), these two brances combine into a
 single mode.  }
\label{como}
\centerline{\psfig{file=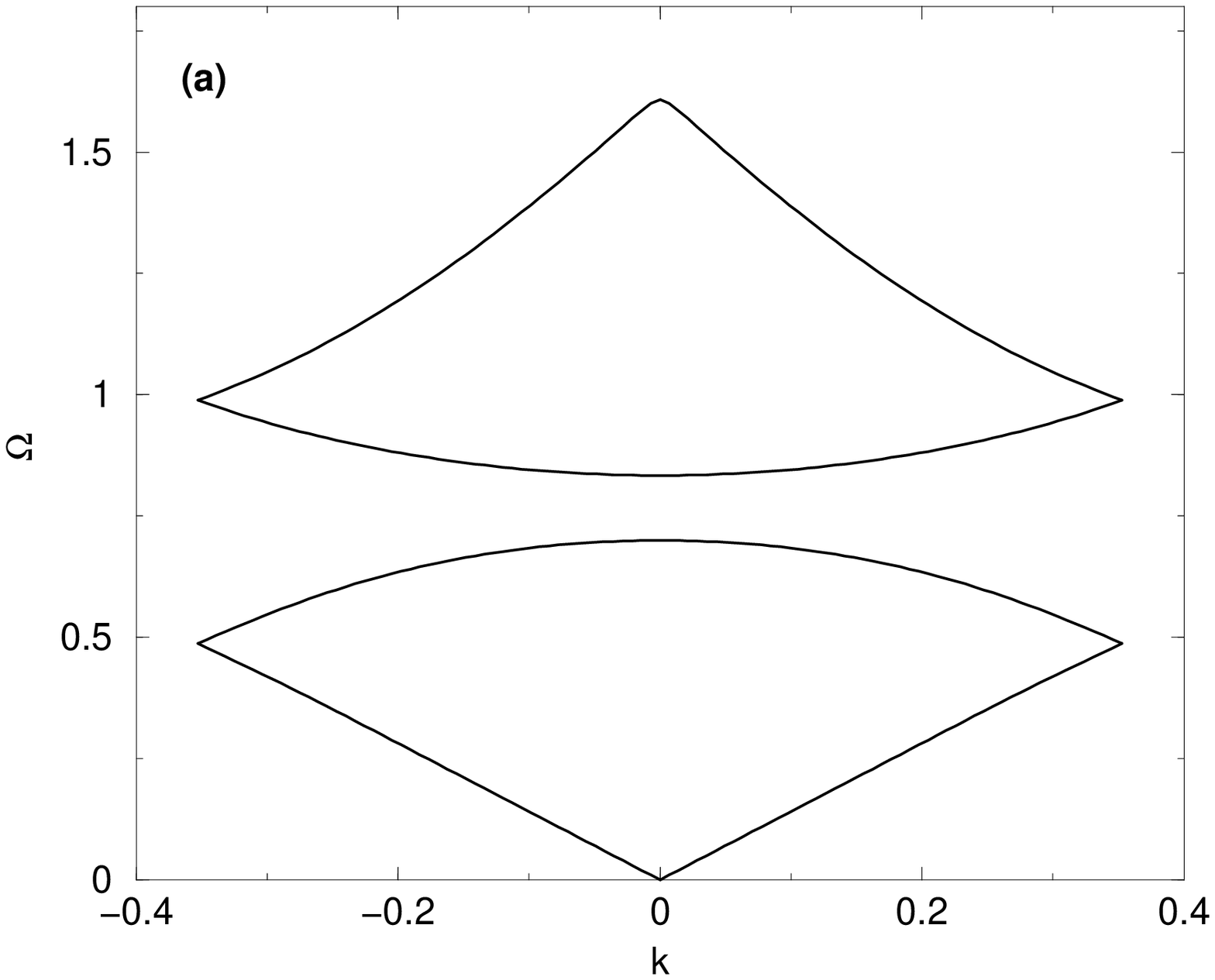,height=9cm,width=10cm,silent=}
\psfig{file=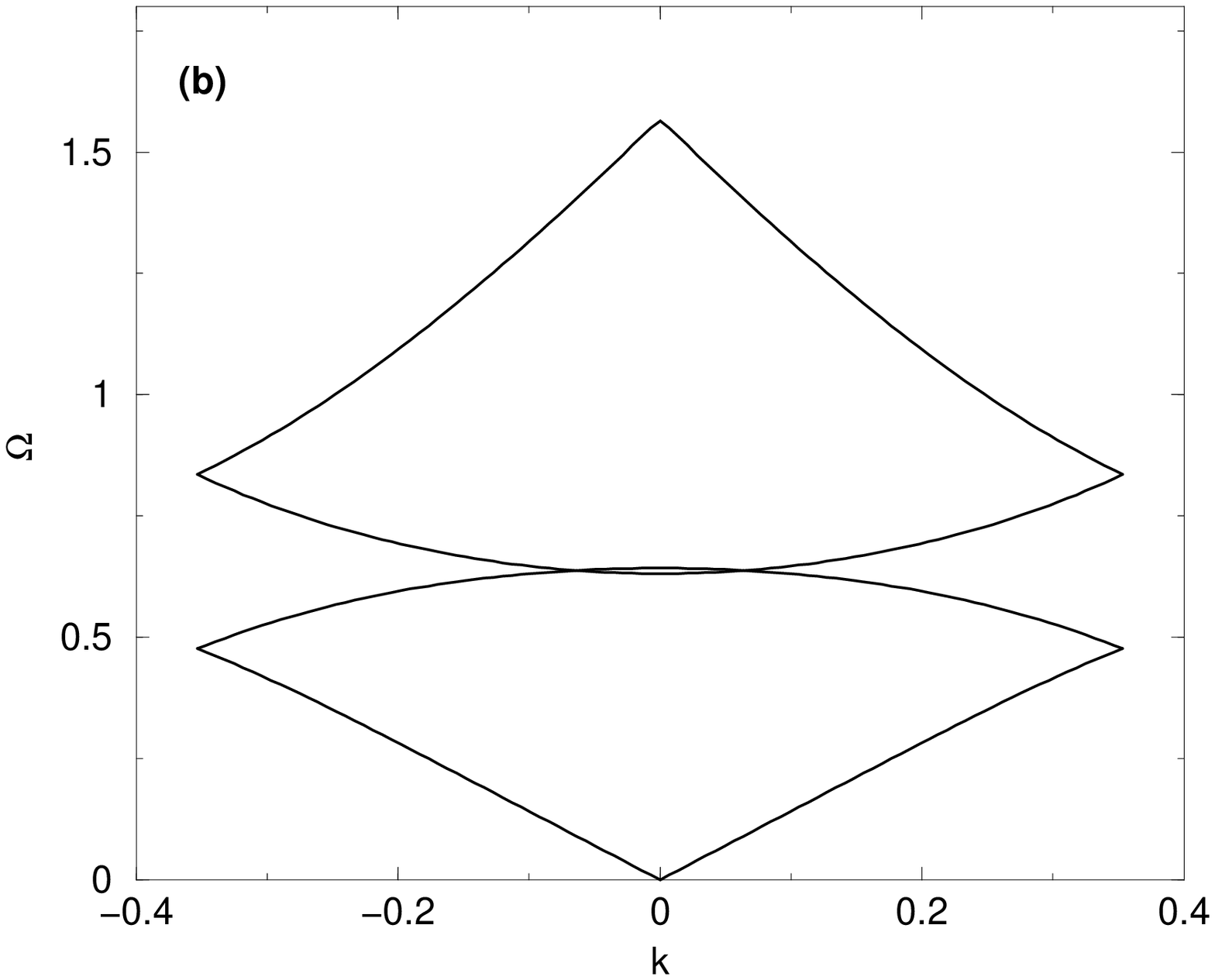,height=9cm,width=10cm,silent=}}
\centerline{\psfig{file=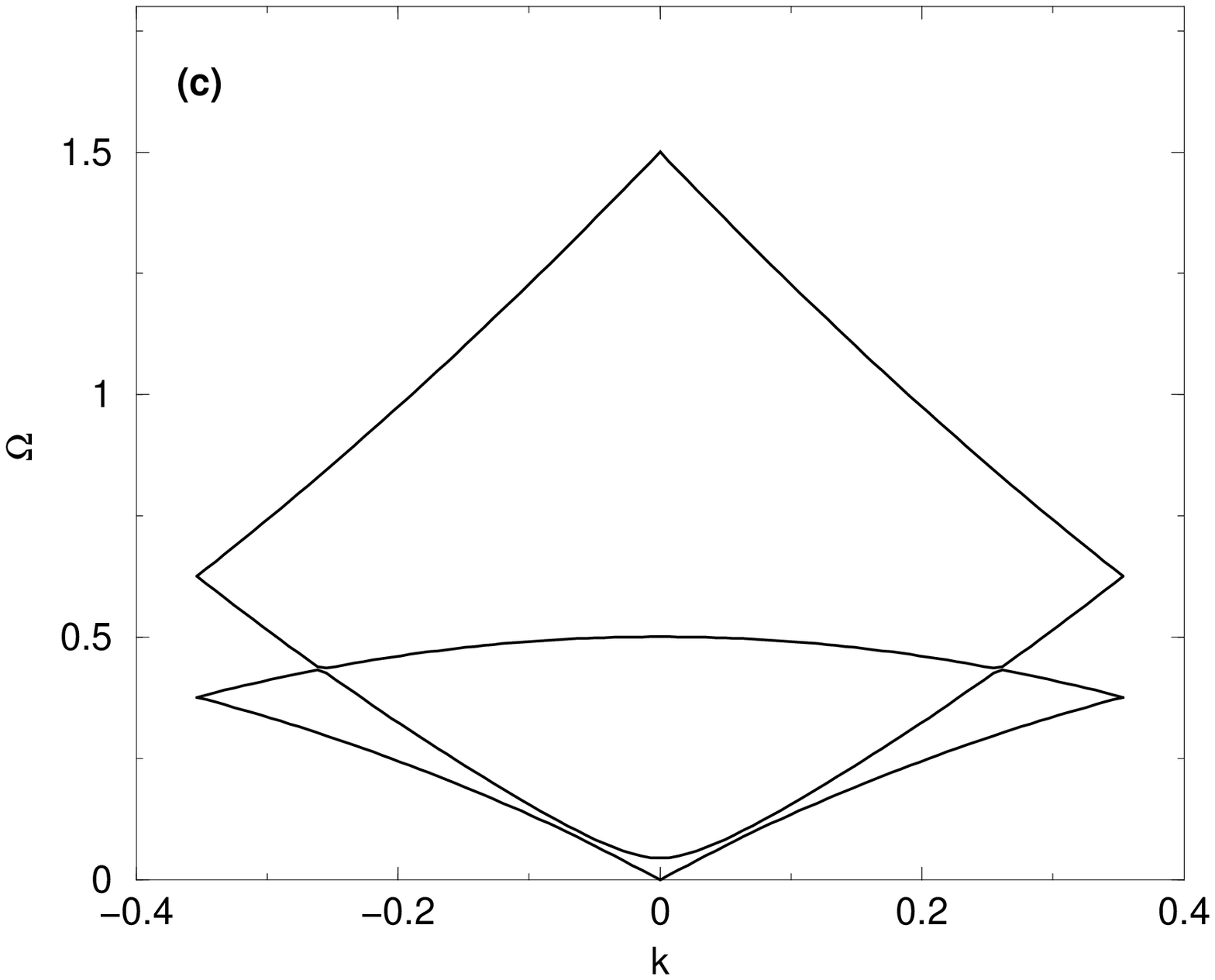,height=9cm,width=10cm,silent=}
\psfig{file=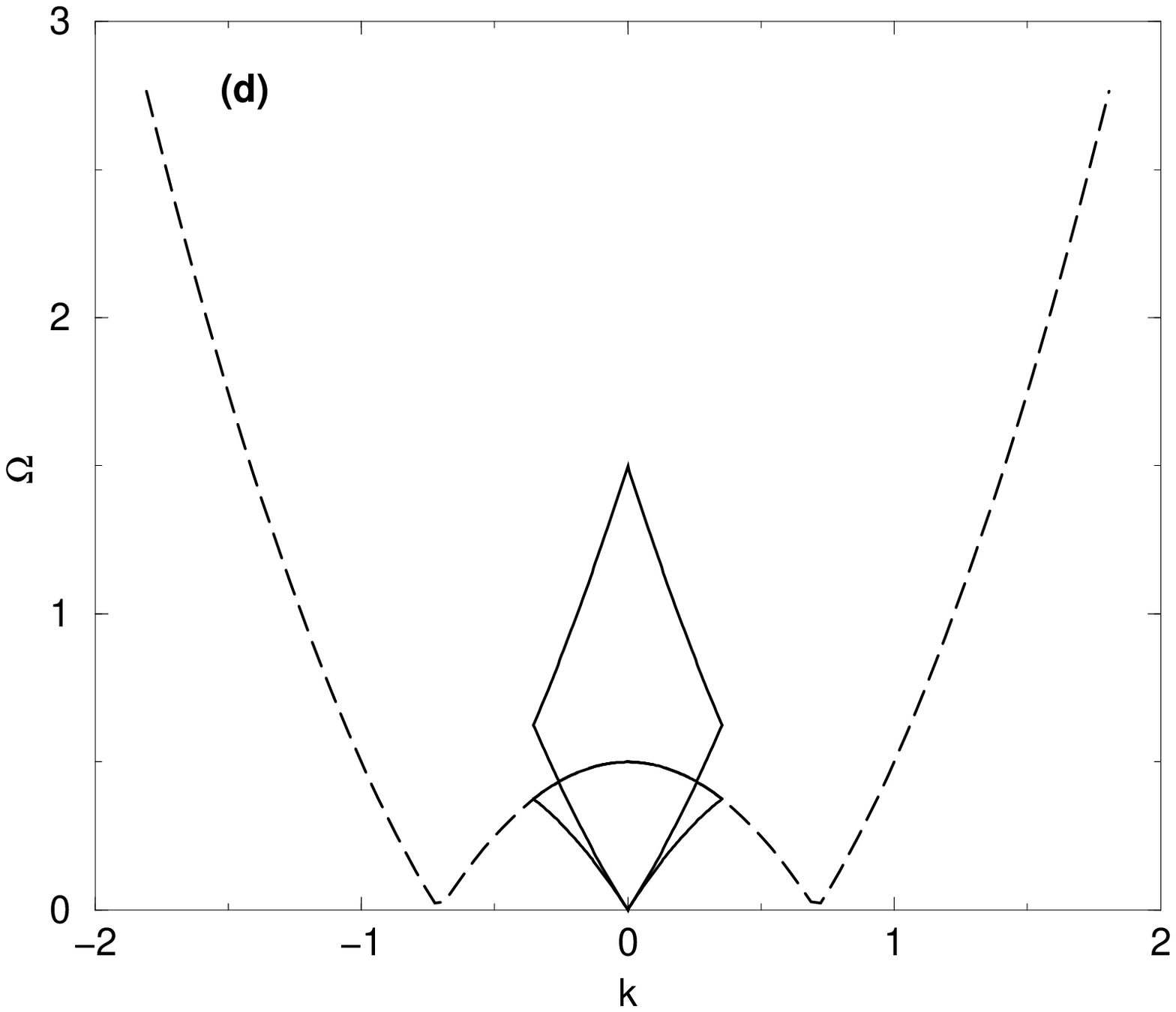,height=9cm,width=10cm,silent=}}
\end{figure}

\begin{figure}
\caption{The dependence of the phase velocity of Goldstone mode on
 $\lambda$, corresponding to the almost sinusoidal configuration $s^{2}$. }
\label{vgsi}
\centerline{\psfig{file=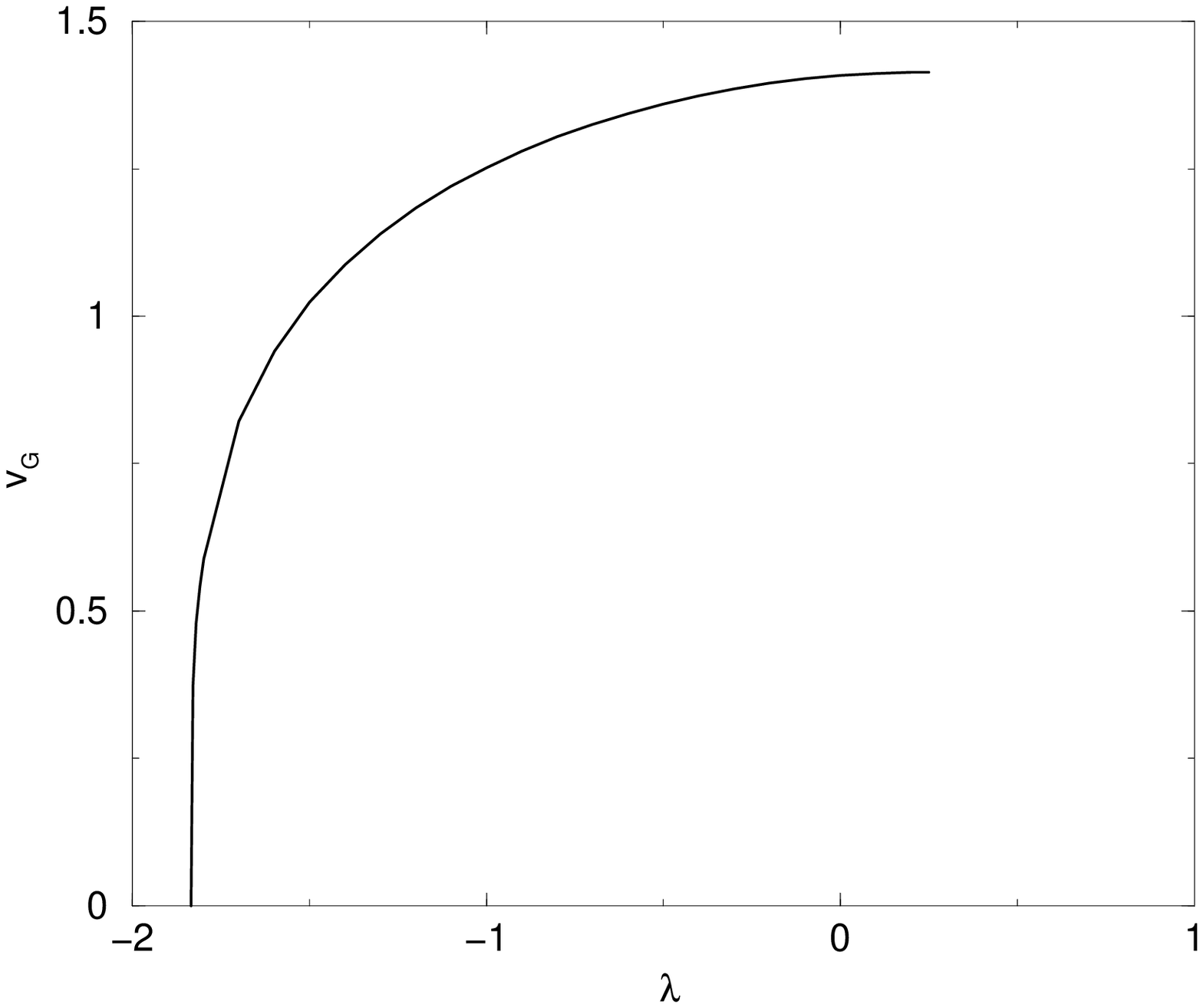,height=10cm,width=11cm,silent=}}
\end{figure}

\begin{figure}
\caption{Four lowest branches of collective modes for the simplest of
 non-sinusoidal state $sd$ for $\lambda=-1$. The branches for all other
 non-sinusoidal configurations are qualitatively same as that for $sd$. }
\label{comod}
\centerline{\psfig{file=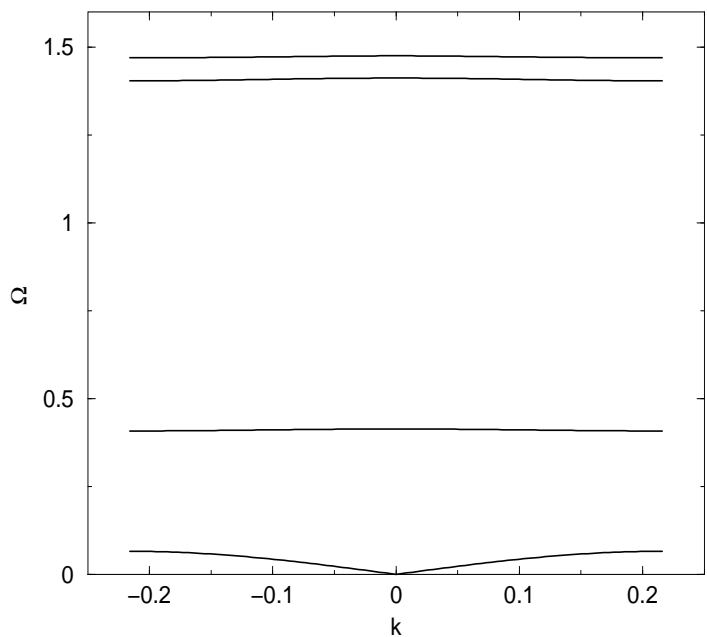,height=10cm,width=11cm,silent=}}
\end{figure}

An illustration of spectra of collective modes for metastable states is shown 
in Fig.~\ref{comod}. We take the state $sd$, chose the value of control 
parameter somewhere in the middle of corresponding region of stability from 
Fig.~\ref{phdi} ($\lambda=-1$), and plot four lowest branches of collective 
modes. Spectra for all other metastable states from Fig.~\ref{phdi} have the 
same qualitative properties, and therefore are not plotted. More specifically, 
for all states, and for all values of $\lambda$ within the respective ranges of 
stability, the subsequent branches are separated by finite gaps, i. e. there is 
no branch overlap, like that obtained for the state $s^2$ [Figs.~\ref{como}a,b].

\begin{figure}
\caption{The dependence of the phase velocity of Goldstone modes on the 
 parameter $\lambda$ for all metastable non-sinusoidal configurations.}
\label{vgall}
\centerline{\psfig{file=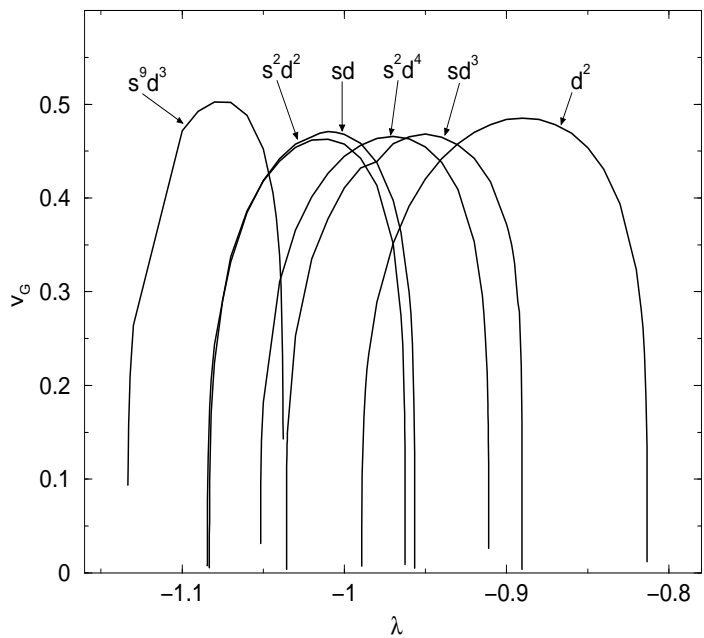,height=10cm,width=11cm,silent=}}
\end{figure}

Furthermore, the lowest branch for all states is the Goldstone mode with the
dispersion $\Omega(k) \approx v_G k$ for $k \rightarrow 0$, and with the 
corresponding phase velocity $v_G$ vanishing for the values of parameter 
$\lambda$ at the edges of stability. The dependence of $v_G$ on $\lambda$ for 
all metastable states from Fig.~\ref{phdi} is shown in Fig.~\ref{vgall}. 
The characteristic scales for these velocities, given by maxima $v_{G,max}$ of 
curves $v_G(\lambda)$ for each metastable state, are situated in the range of 
values ($0.4 - 0.5$ in dimensionless units of Fig.~\ref{vgall}). This is to be 
compared with the maximum value of about $1.4$ of $v_G$ for the configuration 
$u_s(z)$ (Fig.~\ref {vgsi}). In this respect one may recognize a rough tendency 
by which $v_{G,max}$ decreases as the proportion of incommensurate ($s$) 
domains decreases. This decrease is particularly evident as one compares
$s^2$ with $s^9d^3$, and with other configurations from Fig.~1 of 
Ref.~\cite{dbpre}, in which the proportion of commensurate domains $d$ is 
larger and both types of domains become rather short. The decrease
of $v_{G,max}$ is then saturated, i. e. values of $v_{G,max}$ are 
roughly concentrated in the narrow range $0.47 - 0.48$. 

The above tendency can be plausibly interpreted along the lines from the 
preceding subsection. The metastable states are in fact domain trains, built as 
successions of segments with local sinusoidal ($u_s$) and commensurate ($u_c$) 
orderings. As was already stated, it is plausible to associate to a commensurate
segment a Goldstone mode with the vanishing frequency (and the vanishing 
velocity as well). The total Goldstone mode, which is some hybrid of these 
vanishing contributions and the contributions from the local sinusoidal 
ordering, tends to be softer and softer as the train has more and more 
commensurate domains. As a consequence, the velocity $v_{G,max}$ gradually 
decreases as the proportion of commensurate segments in metastable states 
increases.    

\section{Conclusions}

The results presented in Sec.5 show that the spectra of collective 
excitations for all periodic states, stable and metastable, from the phase 
diagram of the model~(\ref{fe1},\ref{fe2}) [Fig.~\ref{phdi}] have Goldstone
branches with a linear dispersion $\Omega = v_G k$ in the long wavelength 
limit. Thus, although these spectra belong to the nonintegrable model, they 
have standard characteristics that essentially follow from the absence of an 
explicit $x$-dependence of free energy density in Eq.~(\ref{fe1}). The 
latter property of the free energy in turn ensures the translational
degeneracy of all solutions of EL equation~(\ref{el}), including those 
participating in the phase diagram. In this respect the present spectrum 
does not differ qualitatively from those of integrable models with the same 
property.

The fact that the chaotic content of the phase space for nonintegrable models,
like for that defined by Eq.~(\ref{fe1})~\cite{dbpra,dbpre} or for other 
examples~\cite{bjbar,lbpl,lbprb}, does not have as substantial impact on the 
spectrum of collective excitations as it has on the thermodynamic phase 
diagram, can be interpreted in the following way. The states from the phase 
diagram belong to the subset of solutions of EL equation defined by conditions 
like Eq.~(\ref{condA}). They are localized in the orbitally unstable chaotic 
layers which cover the phase space, have the measure zero in this space, and 
are mutually separated by topological barriers with characteristic heights 
given by averaged free energies of these layers~\cite{bjbar,kawa}. These 
barriers do not allow for smooth changes from one state to others, and as such 
represent an intrinsic mechanism for frequently observed phenomena like memory 
effects and thermal hysteresis, as discussed in detail in Ref.\cite{dbpre}. In 
general, nonintegrable free energy functionals have more complex phase diagrams
than integrable ones.

On the other hand, collective modes belong to another space of states, denser 
than the phase space, i. e. to that defined by the second order variational 
procedure and the corresponding eigenvalue problem~(\ref{eta}). All states in 
this space are realizable as thermodynamic fluctuations. They have usual 
properties of double periodic linear systems, although the corresponding Bloch 
functions $\Psi_k(z)$ in Eq.~(\ref{bloch}) may be far from a simple sinusoidal 
form. These properties are not essentially dependent on the level of 
integrability of free energy functional.

In order to resolve the eigenvalue problem~(\ref{eta}) for the model~(\ref{fe1},
\ref{fe2}), we formulate here a method based on the general Floquet-Bloch 
formalism, applicable to any IC system showing stable multiharmonic (i.e. 
non-sinusoidal) periodic ordering(s). Beside being a basis for the numerical 
calculations of eigenvalues and eigenfunctions (Sec. 5), this approach clearly 
indicates that for more complex models and orderings the traditional notions of 
phasons and amplitudons are not appropriate. In 
particular, it was often claimed that, being an expansion in terms of real 
order parameter, the functional ~(\ref{fe1}) itself is insufficient for the
stabilization of modulated states in the systems of II class, since  
incommensurate states, in particular those with soliton lattice like
modulations, should have to be described by at least a two-dimensional order
parameter \cite{ls,aramburu,san,ss}. Also, the absence of phase variable 
in  Eq.(\ref{fe1}) caused a belief \cite{mashiyama} that the states which 
emerge from this functional do not 
have an acoustic (phason-like) collective mode. However, while the previous 
study \cite{dbpre} led to the conclusion that almost sinusoidal and highly
non-sinusoidal configurations are among (meta)stable states of the 
model ~(\ref{fe1}) (as is seen in Fig.\ref{phdi}),
the present analysis shows that Goldstone modes are well defined for all these
configurations. From the other side, all dispersive modes for the homogeneous 
($u=$const.) states are massive, i.e. have finite gaps. 

The gap of the lowest such mode tends to zero at continuous (2nd order) 
phase transitions from one homogeneous state to another, or to some periodic 
ordering. The examples are the 
lines ($c>0, a=0$) and ($c<0, \lambda \equiv ad/c^2= 1/4$), 
representing the transitions from the disordered state to the commensurate and 
incommensurate states respectively. As is shown in Fig.~\ref{como},
the situation is qualitatively different at the transition from
the disordered state to the incommensurate, almost sinusoidal, one.
The reason is the specific behavior of Goldstone mode in the 
incommensurate state. By approaching the transition from the incommensurate 
side the phase velocity of this mode, $v_G$, remains finite, while, as is shown
elsewhere~\cite{dbldi}, its oscillatory strength tends to zero. 
In fact, the above behavior of Goldstone mode for the state $s^2$ at the
second order transition to the disordered state is exceptional. Namely, the  
Goldstone modes in the (meta)stable states behave critically at the edges of 
stabilities for these states, including the lower edge of state $s^2$ at
$\lambda=-1.835$. At these edges the phase velocities $v_G$ vanish. 
All these specific properties of collective modes, particularly of the most
interesting Goldstone modes, are expected to be directly experimentally
observable in X-ray and neutron scatterings, as well as in optical and similar
measurements. The particular discussion on the role of these collective
modes in the dielectric response, and the comparison with measurements
on some materials of the II class, is given in Ref.\cite{dbldi}.  

Finally, we comment on the general property of  Goldstone modes for metastable
periodic states by which they become softer and softer as the period of these 
states increases. This tendency, shown in Fig.~\ref{vgall}, has its origin in 
the elastic nature of Goldstone modes in the long wavelength limit. More 
specifically, as the segments of local sinusoidal order become more and more 
dilute in the underlying commensurate background, the slight variations in 
their mutual distances cost less and less energy, i. e. the corresponding 
effective elastic constant decreases. In this interpretation, which holds for 
dilute soliton lattices as well, the commensurate ordering is by assumption 
perfectly elastic, i. e. the notion of relative distance has no sense since the 
lattice discreteness is neglected. The only possible deformations are those 
invoking the variations of amplitude, and resulting in the massive collective 
modes. The lattice discreteness introduces, through an "external" potential of 
Peierls-Nabarro type, the finite stiffness of the local commensurate ordering, 
or even opens the gap in the Goldstone mode for dilute incommensurate states at 
the transition by breaking of analyticity \cite{aubry,baesens,lorenzo}. 

\bigskip

{\bf ACKNOWLEDGMENTS}

\bigskip

The work was supported by the Ministry of Science and Technology of 
the Republic of Croatia through project No. 119201.

\appendix

\section{}

The procedure from Sec. 3 takes into account, after relaxing boundary conditions
(\ref{bv}), {\em all infinitesimal} variations of the order parameter $u(z)$.
This generalization includes some thermodynamic variations, like e. g. those,  
specified by the scaling $u(z)\rightarrow su(z)$ with $s \rightarrow 1$, 
responsible for the condition obeyed by $u(z)$ at boundaries $z=0$ and $z=L$ 
(condition B in Ref.~\cite{dbprl}). However, by this procedure the analysis of 
thermodynamic stability is still not completed, since there remain variations 
which invoke infinitesimal relative changes in the configuration $u(z)$, but 
are not infinitesimal at the absolute scale. An example is the scaling 
\begin{equation}     
u(z)\longrightarrow u[(1+\epsilon)z],\,\,\,\,\,\,\,\,\,\,
\epsilon \rightarrow 0,
\label{epsilon}
\end{equation}
which leads to the condition~(\ref{condA}). The variation that corresponds 
to this scaling is not infinitesimal. Indeed, after the transformation 
$(1+\epsilon)z \rightarrow z$ in the integral~(\ref{fe2}), it follows that 
this variation behaves as $z$ and therefore does not fulfill the  
criterion of infinitesimality specified in Sec. 3. Thus the above procedure 
has to be enlarged by including the expansion of the free energy with respect 
to $\epsilon$ up to the quadratic terms. While the requirement that the linear 
term vanishes gives the condition~(\ref{condA}), the second order variation 
reads
\begin{eqnarray}
\delta^{2}f & \equiv & f\left[u((1 + \epsilon)z) + \eta\right] - f[u(z)] =
 \nonumber \\ 
& = & \frac{1}{L}\int_{0}^{L} dz \left[\eta(z) {\cal D} \eta(z) + 
2(u'(z))^2 \epsilon^2 + 4(2u''''(z) + u''(z))\eta(z) \epsilon \right],
\label{fevare}
\end{eqnarray}
i.e. the expression~(\ref{fevar}) is extended  by the term quadratic in 
$\epsilon$, and the term representing the bilinear coupling between $\eta(z)$ 
and $\epsilon$.
 
The previous analysis~\cite{dbpra,dbpre} of the model~(\ref{fe1},\ref{fe2}) 
led to the conclusion that all solutions $u(z)$ of the EL equation~(\ref{el}) 
that participate in the thermodynamic phase diagram as stable 
or metastable configurations, are simple periodic. The analysis in Sec.4
shows that the corresponding eigenfunctions $\eta_{\Lambda}(z)$ of the 
problem~(\ref{eta}) are then double periodic. This means that for periodic 
extrema $u(z)$ the bilinear coupling in the expression~(\ref{fevare}) vanishes, 
i. e. the fluctuations in $\epsilon$ are decoupled from $\eta(z)$ fluctuations. 
The remaining $\epsilon^2$ term is positively definite, i. e. all periodic
configurations satisfying the EL equation~(\ref{el}) and the
condition~(\ref{condA}) are also stable with respect to the variation defined 
by the scaling~(\ref{epsilon}), irrespectively to the value of the control 
parameter $\lambda$ figuring in the functional~(\ref{fe2}).


\newpage

\newpage

\newpage

\end{document}